\documentclass[aps,prl,amsmath,amssymb,
reprint,superscriptaddress,longbibliography]{revtex4-2}
\usepackage{graphicx}
\usepackage{dcolumn}
\usepackage{physics}
\usepackage{xspace}
\usepackage{bm}
\usepackage[normalem]{ulem}

\begin{document}

\preprint{APS/123-QED}

\title{
  Inferring Coupled Stuart-Landau Equations from Waveforms
}

\author{Yuki Araya}
\email[]{araya.yuuki.work@gmail.com}
\affiliation{Department of Physics, Chiba University, Chiba 263-8522, Japan}
\author{Hiroaki Ito}
\email[]{ito@chiba-u.jp}
\affiliation{Department of Physics, Chiba University, Chiba 263-8522, Japan}
\author{Hiroshi Kori}
\email[]{kori@k.u-tokyo.ac.jp}
\affiliation{Graduate School of Frontier Sciences,
The University of Tokyo, Chiba 277-8561, Japan}
\author{Hiroyuki Kitahata}
\email[]{kitahata@chiba-u.jp}
\affiliation{Department of Physics, Chiba University, Chiba 263-8522, Japan}

\date{\today}

\begin{abstract} 
We present a physics-informed, data-driven framework for inferring phase-amplitude equations of coupled limit-cycle oscillators from waveform measurements.
Guided by normal-form theory near a supercritical Hopf bifurcation, we infer a near-identity transformation from two observables of an isolated oscillator to a complex-amplitude coordinate, together with the intrinsic Stuart-Landau dynamics.
Applying this transformation to coupled oscillators then enables estimation of the leading-order linear coupling terms.
Using coupled van der Pol oscillators as an analytical benchmark, we demonstrate accurate inference under system noise and quantify the validity range with respect to the Hopf bifurcation parameter and coupling strength.
Moreover, a model inferred from one informative trajectory at each coupling condition reproduces synchronized states and their basin structure for initial conditions not used in the inference.
Applied to a high-dimensional hydrodynamic system, the inferred model captures bistability between in-phase and anti-phase synchronization and identifies the loss of anti-phase synchrony as a Neimark-Sacker bifurcation.
Our framework provides a practical route from experimentally accessible waveforms to an interpretable normal-form model for stability and bifurcation analysis.
\end{abstract}

\maketitle

{\em Introduction}---Synchronization, the phenomenon in which self-sustaining oscillators adjust their rhythms through interaction, appears in various systems~\cite{winfree01,kuramoto84,pikovsky01,ashwin2016mathematical}.
Examples include mechanical systems~\cite{martens2013chimera,goldsztein2021synchronization,kato2024weakly}, chemical systems~\cite{kiss02,kiss07,toiya2010synchronization}, cell populations~\cite{glass01} such as cardiac cells~\cite{clay79} and clock cells~\cite{yamaguchi03}, and fluid interaction systems of flagella and cilia~\cite{uchida2010synchronization,golestanian2011hydrodynamic,brumley2014flagellar}.
Synchronization is essential for the functioning of many artificial and biological systems.

\begin{figure}[tb]
  \includegraphics[scale=0.94]{"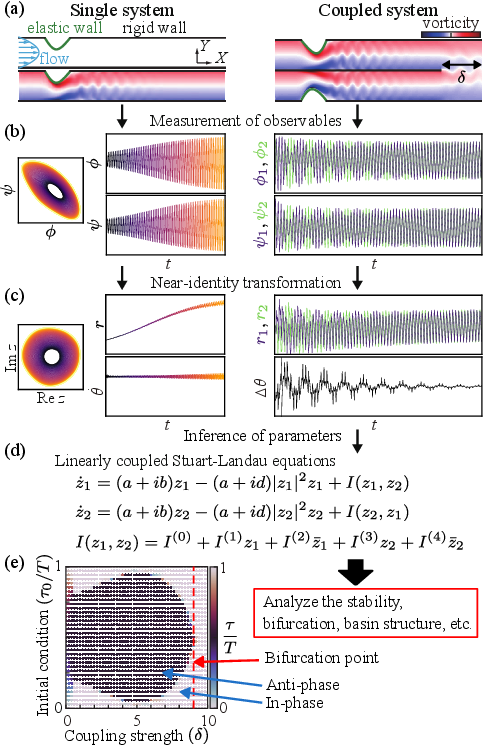"}
  \caption{
      Schematic overview of the proposed approach with data from a hydrodynamic system used for demonstration.
    (a) Snapshot of the hydrodynamic system.
    The system shows bistability between in-phase and anti-phase synchronization (see details in (e)).
    (b) Example time series $\phi(t)$ and $\psi(t)$, which can be two independent waveform observables    from the system.
    (c) Result of the optimal variable transformation, which maps the original variables $(\phi,\psi)$ to the rotation-symmetric coordinates $z=r e^{i \theta}$.
    (d) The estimated model, in which the transformed variables in (c) are expected to follow in a good approximation.
    Colors indicate the time course in the panels (b) and (d).
    As will be shown in Fig.~\ref{fig:3}, analysis of the estimated model in (d) reveals that the loss of the anti-phase synchronized state occurs via a Neimark-Sacker bifurcation.
    (e) Color map that represents the final phase difference $\tau/T$ after transients in the plane spanned by the initial phase difference $\tau^{(0)}/T$ between the oscillators (vertical axis) and the coupling strength $\delta$ (horizontal axis), the system exhibits bistability between in-phase and anti-phase synchronization for small $\delta$, while the anti-phase state disappears near $\delta \simeq 9$.
  \label{fig:1}}
\end{figure}

Most oscillator populations, with some exceptions, are inherently high-dimensional systems.
Nevertheless, low-dimensional dynamics often play a pivotal role in shaping collective behavior, even within such complex systems.
As a result, reducing these systems to low-dimensional representations has been fundamental to the universal and unified understanding of synchronization phenomena.
A prominent example is the phase reduction method~\cite{kuramoto84,ashwin2016mathematical,Nakao2016ContPhys}.
This approach models each oscillator using a single variable, the phase, and provides a reliable approximation of the system's dynamics when interactions and perturbations are sufficiently weak.
Beyond theoretical insights, phase equations have been successfully applied to the design and control of synchronization patterns in complex systems~\cite{kiss07,kori08}.
In recent years, considerable efforts have been made to infer phase equations directly from empirical data obtained from coupled oscillator systems~\cite{kiss07,tokuda2007inferring,blaha2011reconstruction,arai2022extracting,rosenblum2023inferring,matsuki2025network}.
These data-driven approaches enable the extraction of key information, such as coupling structures, interaction strengths, noise levels, and network topology.

However, the dynamics of oscillatory systems cannot be fully captured by phase equations alone.
Many phenomena critically depend on amplitude variations, including continuous transitions from damped oscillations to limit cycles via Hopf bifurcations, destabilization of limit cycles through period-doubling or Neimark-Sacker bifurcations, and various routes to chaos~\cite{kuznetsov2004elements}.
To understand such phenomena, low-dimensional models known as complex amplitude equations or phase-amplitude equations are particularly useful.
These equations can be systematically derived when the governing equations of the system are known~\cite{kuramoto84,Gengel2021,Wilson2019}.
Recent data-driven equation-discovery methods, such as sparse identification of nonlinear dynamics (SINDy), infer parsimonious governing equations by selecting active terms from a prescribed library of candidate functions~\cite{Brunton2016}.
However, their performance generally depends on the choice of state coordinates in which the dynamics admit a simple representation.
This coordinate dependence has motivated approaches that jointly learn latent coordinates and governing equations using autoencoders and sparse identification~\cite{Champion2019}.

In this study, we address this coordinate problem in a physics-informed manner and propose a methodology for estimating phase-amplitude equations of coupled oscillators near a supercritical Hopf bifurcation.
In conventional normal-form analysis, a near-identity transformation is derived from known governing equations to obtain a complex-amplitude coordinate. 
Here, we infer this transformation from waveform measurements and estimate the Stuart-Landau dynamics in the reconstructed coordinate. 
This provides a direct route from measured waveforms to a normal-form model without first identifying detailed governing equations in the measurement coordinates.
The resulting physics-informed coordinate reconstruction reduces the model-selection problem and facilitates inference from relatively short and noisy waveform data.
The inferred model can be used to examine dynamics beyond the trajectories used for inference, including synchronized states, their basin structure, stability, and bifurcations.

Figure~\ref{fig:1} provides an overview of the two-stage inference procedure and its application. First, two waveform observables, $\phi(t)$ and $\psi(t)$, from an isolated oscillator are used to infer the near-identity transformation and the intrinsic Stuart-Landau parameters.
Second, the inferred transformation is applied to waveform observables, $\phi_i(t)$ and $\psi_i(t)$, from coupled oscillators to estimate the leading-order linear interaction terms.
Using coupled VDP oscillators as an analytical benchmark, we validate the inferred parameters and dynamical properties over finite ranges of coupling strength and distance from the Hopf bifurcation and under additive white noise.
A model inferred from a single informative trajectory also reproduces synchronized states and their basin structure for initial conditions not used for inference.
Applied to a high-dimensional hydrodynamic system, the inferred model captures the bistability between in-phase and anti-phase synchrony
and identifies the stability loss of anti-phase synchrony
as a Neimark-Sacker bifurcation.

{\em Methodology}---We first infer the coordinate transformation and the intrinsic Stuart-Landau dynamics from isolated-oscillator data.
Near a supercritical Hopf bifurcation, two independent observables $\phi(t)$ and $\psi(t)$ can be parametrized, to the order considered here, by a complex amplitude $z(t)$ and its complex conjugate $\bar{z}(t)$, where
\begin{align}
    \dot z=(a+ib)z-(c+id)|z|^2z, \label{quad:1}
\end{align}
with real parameters $a$, $b$, $c$, and $d$.
By rescaling $z$, we set $c=a>0$ without loss of generality, so that the amplitude of the stable limit cycle is unity.
Its angular frequency is then $b-d$, which is determined from the period of the asymptotic waveform.

The central step of our method is to reconstruct from the measured variables $(\phi,\psi)$ a complex-amplitude coordinate adapted to the Hopf normal form.
Guided by normal-form theory~\cite{Guckenheimer,Wiggins2003}, we represent the coordinate transformation as a linear mapping
\begin{align}
    w=q_\phi\phi+q_\psi\psi-C^\qty(0,0), \label{linear}
\end{align}
followed by the near-identity transformation
\begin{align}
    w=z+h_2(z,\bar{z})+h_3(z,\bar{z})+\order{4}, \label{quad:4}
\end{align}
where $h_2$ and $h_3$ are second- and third-order polynomials, respectively, and $\order{4}$ denotes terms $z^n\bar{z}^m$ with $n+m\geq4$.
Their explicit forms and the derivation of the transformation are provided in Sec.~\ref{SM:sec1_coordinate} of the Supplemental Material (SM)~\cite{SM}.

Most transformation coefficients are first obtained from the Fourier components of the asymptotic waveforms.
To infer the remaining transformation coefficients together with the intrinsic Stuart-Landau parameters, we invert Eq.~\eqref{quad:4} and reconstruct a complex amplitude $\hat{z}(t)$ from the measured waveforms.
These parameters are then determined by matching $\hat{z}(t)$ to the analytical solution of Eq.~\eqref{quad:1} over a transient time window.

We next infer the coupling terms from waveform data for coupled oscillators.
In the reconstructed complex-amplitude coordinates, we consider
\begin{align}
 \dot{z}_k =& G_k(z_k,z_\ell) = (a+ib)z_k-(a+id)|z_k|^2 z_k \nonumber \\
 &+I^\qty(0) +I^\qty(1)z_k+I^\qty(2)\bar{z}_k+I^\qty(3)z_\ell+I^\qty(4)\bar{z}_\ell,
\label{coupledSL}
\end{align}
where $(k,\ell)=(1,2)$ or $(2,1)$ and $I^\qty(0),\ldots,I^\qty(4)$ are complex coupling coefficients.
We retain the leading-order linear coupling terms, which are expected to provide the dominant contribution in the small-amplitude regime near the Hopf bifurcation.

Using the transformation inferred from the isolated oscillator, we reconstruct $\hat{z}_k(t)$ from the measured waveforms of the coupled system.
The coupling coefficients are inferred by matching the reconstructed time derivatives, evaluated using a second-order central difference, to the vector field $G_k(\hat{z}_k,\hat{z}_\ell)$.

In the present implementation, the parameters to be fitted in both the isolated- and coupled-oscillator inference are estimated using differential evolution.
The objective functions and numerical implementation are described in Sec.~\ref{SM:Inference_detail}, and the system-specific settings are given in Sec.~\ref{SM:sec_system_settings} of SM~\cite{SM}.
As demonstrated below, the inferred parameters are stable over finite ranges of data windows in both inference stages, over a finite range of coupling strengths in the coupled system, and under the investigated levels of additive white noise.

{\em Verification}---We verify our proposed method using a low-dimensional oscillator model.
Specifically, we employ the coupled VDP oscillators expressed as
\begin{align}
  \ddot x_k=(\mu-x_k^2)\dot x_k-x_k-I_x(x_k-x_\ell)-I_v(\dot x_k-\dot x_\ell),
 \label{vdp}
\end{align}
where $x_{k,\ell}$ with $(k,\ell)=(1,2)$ and $(2,1)$ are real variables; $\mu$, $I_x$, and $I_v$ are real constants.
For this model, as detailed in Sec.~\ref{SM:VDP_analysis} of SM~\cite{SM}, we obtain the following expressions to the lowest order of $\mu$: 
$a=\mu/2$, $b=1$, $d=0$, $I^\qty(0)=0$, $I^\qty(3)=-I^\qty(1)=I_v/2-i(2I_x+\mu I_v)/4$, $I^\qty(4)=-I^\qty(2)$, and $\qty|I^\qty(2)|=\qty|I^\qty(1)|$. 
The arguments of $I^\qty(2)$ and $I^\qty(4)$ are uniquely determined if the transformation between $z_k$ and $(x_k,\dot{x}_k)$ is fixed (see Sec.~\ref{SM:VDP_analysis} of SM~\cite{SM}).
Stability boundaries in the $(I_v,I_x)$ plane are obtained using these expressions (see Sec.~\ref{SM:cSL_stability} of SM~\cite{SM}).
There is a region of bistability of in-phase and anti-phase synchrony, which is surrounded by the pitchfork and the Neimark-Sacker bifurcation curves (see Fig.~\ref{fig:4} in SM~\cite{SM}).
\begin{figure}[tb]
  \includegraphics{"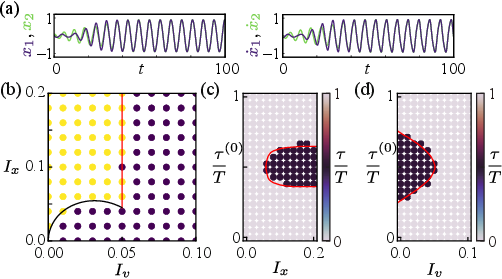"}
  \caption{
Inference from single transient trajectories of the coupled van der Pol (VDP) oscillators at $\mu=0.2$.
(a) Example of a trajectory used for inference at $I_x=0.16$ and $I_v=0.04$. The oscillators were initialized with amplitudes substantially smaller than the isolated limit-cycle amplitude and with the initial phase difference $\tau^{(0)}/T=0.25$.
For each pair $(I_x,I_v)$, a coupled Stuart-Landau model was inferred independently from only one trajectory generated at that parameter pair with the same initial phase difference.
(b) Synchronization diagram in the $(I_v,I_x)$ plane obtained from the inferred models.
Yellow circles indicate bistability between the in-phase (IP) and anti-phase (AP) synchronized states, whereas purple circles indicate monostability of the IP state.
The black and red solid curves are the pitchfork [Eq.~\eqref{full_PF}] and Neimark-Sacker [Eq.~\eqref{NS}] bifurcation boundaries, respectively, obtained in Sec.~\ref{SM:cSL_stability} of SM~\cite{SM}.
(c)--(d) Basin structures obtained from the inferred models as functions of the initial phase difference $\tau^{(0)}/T$ and (c) $I_x$ at fixed $I_v=0.03$, or (d) $I_v$ at fixed $I_x=0.16$.
The color indicates the final phase difference $\tau/T$ after transients.
At each coupling condition, only the trajectory with $\tau^{(0)}/T=0.25$ was used for inference; all other initial phase differences shown in the basin maps were used only for validation.
The red solid curves in (c) and (d) indicate the basin boundaries obtained directly from numerical simulations of the original coupled VDP equations.
}  
  \label{fig:2}
\end{figure}

We apply our method to Eq.~\eqref{vdp}, using $x_k(t)$ and $\dot{x}_k(t)$ as observables.
For the isolated oscillator, we use a transient initialized away from the stable limit cycle. 
As a baseline test for the coupled system, we first infer the coupling coefficients simultaneously from 21 transient trajectories initialized at equally spaced phase differences $\tau^{(0)}/T$.
The fitting windows are selected from broad regions in which the inferred parameters are insensitive to moderate variations in the start time $t_{\rm s}$ and duration $t_{\rm w}$
specifying the data segment used for inference. Details are provided in Sec.~\ref{SM:sec_system_settings} of SM~\cite{SM}.

Using this trajectory ensemble, we quantified the validity range with respect to coupling strength and distance from the Hopf bifurcation.
Although some inferred coefficients deviate from their leading-order analytical values, the oscillation frequencies and dominant Floquet exponents remain accurately reproduced over the investigated ranges.
The complete results are provided in Sec.~\ref{SM:sec_validity} of SM~\cite{SM}.

We next ask whether the reduced model can be inferred from only one transient trajectory at each coupling condition.
For every pair $(I_x,I_v)$, we independently infer a coupled SL model from a single trajectory with amplitudes substantially smaller than the isolated limit-cycle amplitude.
Figure~\ref{fig:2}(a) shows an example at $I_x=0.16$ and $I_v=0.04$.
The pronounced amplitude-relaxation dynamics cause each fitting trajectory to sample a dynamically informative region of phase-amplitude state space.

The independently inferred models reproduce the organization of the IP and AP synchronized states in the $(I_v,I_x)$ plane [Fig.~\ref{fig:2}(b)].
They also closely reproduce the basin structures of the original coupled VDP equations as either $I_x$ or $I_v$ is varied [Figs.~\ref{fig:2}(c) and \ref{fig:2}(d)].
At each coupling condition, the initial phase differences other than $\tau^{(0)}/T=0.25$ were not used for parameter inference.
The superimposed red curves show the basin boundaries computed directly from the original VDP equations.
Thus, one informative trajectory at each coupling condition can be sufficient to infer a reduced vector field that reproduces multistable dynamics for initial conditions not used in fitting.

Additional limited-trajectory tests, extension to three coupled VDP oscillators, and robustness under additive white noise are documented in Secs.~\ref{SM:limit_number_trajectory}, \ref{SM:three_coupled}, and \ref{SM:noisy_system} of SM~\cite{SM}, respectively.

{\em Application}---Next, we adopted the proposed method for two coupled oscillatory flows in two collapsible channels merging into a single channel (See Fig.~\ref{fig:1}(a)).
When the softness of the elastic wall exceeds a critical threshold, oscillatory flow emerges in each channel via a supercritical Hopf bifurcation. 
In this study, we focus on the conditions slightly above this bifurcation point.
The strength of the coupling between two channels is controlled by the distance $\delta$ between the merging point and the outlet~\cite{Araya2024}.
For $\delta=0$, the channels are independent. The details of hydrodynamic numerical simulation are shown in Sec.~\ref{SM:Fluid_system_Floquet} of SM~\cite{SM}.
As shown in Fig.~\ref{fig:1}(e), the oscillatory flows can synchronize in-phase or anti-phase depending on the value of $\delta$ and the initial phase difference.
We observe that the in-phase synchrony is possible for all the range of $\delta$ under consideration, whereas the anti-phase synchrony does not occur for $\delta>\delta_{\rm c}$, where $\delta_{\rm c} \simeq 9.0$ in the employed setup. 
\begin{figure}[tb]
 \includegraphics{"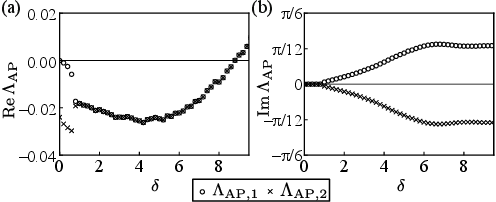"}
  \caption{
  Stability analysis of the anti-phase synchronized periodic solution, based on the Stuart-Landau oscillator system estimated from time series data of a fluid system.
(a) Real and (b) imaginary parts of the Floquet exponents. 
  \label{fig:3}}
\end{figure}

We demonstrate that our inference method enables us to perform the stability analysis and clarify that the anti-phase synchrony is destabilized through a Neimark-Sacker bifurcation at $\delta=\delta_{\rm c}$.
We use the positions $(X, Y)$ of the lowest part of the elastic wall as observables for each oscillator (see Fig.~\ref{fig:1}(a)).
The initial conditions are constructed in the same manner as for the VDP oscillators, with $\tau^{(0)}$ selected from 34 equally spaced points over the interval $[0, T)$, where $T$ denotes the period of the reference periodic solution.
The details of inference are shown also in Sec.~\ref{SM:Inference_detail} of SM~\cite{SM}.

We numerically computed the Floquet exponents of the inferred coupled SL equations including all inferred terms.
As shown in Figs.~\ref{fig:3}(a) and \ref{fig:3}(b), the real parts of a complex-conjugate pair of Floquet exponents cross zero while their imaginary parts remain nonzero, identifying the loss of AP stability as a Neimark-Sacker bifurcation.
The predicted critical value closely agrees with that observed in the original fluid system [Fig.~\ref{fig:1}(e)].
The inferred model also captures the main basin structure when the inference is performed using only two fluid-system trajectories, as detailed in Sec.~\ref{SM:sec_fluid_basin_limited} of SM~\cite{SM}.

{\em Discussion}---The main advantage of the present framework lies in reconstructing a normal-form coordinate from measured waveforms rather than fitting the dynamics directly in measurement-dependent coordinates.
This coordinate reconstruction reduces the model-selection problem and provides a compact phase-amplitude model suitable for stability and bifurcation analyses, including amplitude-dependent instabilities that cannot generally be described by phase-only models. 

The present formulation is limited by the truncated Hopf normal form and the leading-order coupling approximation, whose validity ranges were quantified in the VDP model.
The demonstration for three coupled oscillators provides a first step beyond two-oscillator systems, but scalability to larger networks remains to be established.
Extensions to nonlinear coupling, time-delayed interactions, and higher-order interaction structures are important directions for future work.
Further improvements in the numerical implementation, including numerical differentiation, noise handling, objective-function design, and optimization, may enhance inference from noisy or limited data.

Koopman-based methods provide a complementary framework for constructing phase-amplitude coordinates through eigenfunctions related to isochrons and isostables~\cite{mauroy2013isostables,mauroy2018global}, and have recently been extended to reconstruct phase-amplitude network models directly from observed signals~\cite{yeldesbay2025reconstruction}.
Our approach instead focuses on the vicinity of a supercritical Hopf bifurcation and exploits its normal-form structure to obtain a coupled Stuart-Landau model with a small number of parameters.
Its strength lies not in generality, but in providing a compact, physically structured model whose coefficients can be directly used to analyze the stability and bifurcations of synchronized states.

\begin{acknowledgments}
    {\em Acknowledgments}---We thank Istv{\'a}n Z. Kiss and John L. Hudson for motivating discussions.
    This work was supported by JST SPRING Grant Number JPMJSP2109 (Y.~Araya), by the Cooperative Research Program of ``Network Joint Research Center for Materials and Devices.'' (Nos.~20245003 (Y.~Araya), 20244003, 20254003, and 20264003 (H.~Kitahata)), by JSPS KAKENHI Grants Nos.~JP21K13891 and JP24K06972 (H.~Ito); JP21K12056 and JP25K15258 (H.~Kori); JP24K06978, JP24K22311, JP25K00918, and JP26K00671 (H.~Kitahata) and by MEXT KAKENHI Grant No.~JP26H00384 (H.~Kitahata). This work was also supported by JSPS Core-to-Core Program ``Advanced core-to-core network for the physics of self-organizing active matter (JPJSCCA20230002)'' and MEXT Promotion of Distinctive Joint Usage/Research Center Support Program (JPMXP0724020292).
 
\end{acknowledgments}

\bibliography{manuscript}

\clearpage

\newpage

\onecolumngrid

\renewcommand{\thefigure}{S\arabic{figure}}
\renewcommand{\theequation}{S\arabic{equation}}

\begin{center}
\textbf{\large Supplemental Material: Inferring Coupled Stuart-Landau Equations from Waveforms}
\end{center}

\begin{center}
{\normalsize Yuki Araya, Hiroaki Ito, Hiroshi Kori, Hiroyuki Kitahata}
\end{center}

\setcounter{page}{1}
\setcounter{secnumdepth}{3}
\renewcommand{\theequation}{S\arabic{equation}}
\setcounter{equation}{0}
\setcounter{figure}{0}
\thispagestyle{empty}

\section{Detailed inference methodology \label{SM:Inference_detail}}

This section provides the details of the two-stage inference procedure summarized in the main text.
First, two waveform observables from an isolated oscillator are used to reconstruct a complex-amplitude coordinate and infer the intrinsic Stuart-Landau (SL) dynamics.
Second, the inferred coordinate transformation is applied to coupled-oscillator waveforms, and the coupling coefficients are inferred by matching the reconstructed dynamics to the vector field of the coupled SL equations.

\subsection{Coordinate reconstruction from waveform measurements}
\label{SM:sec1_coordinate}

We express two real observables $\phi$ and $\psi$ of an isolated oscillator in terms of the complex amplitude $z$ and its complex conjugate $\bar z$ as
\begin{align}
 \phi&=\sum_{\substack{n,m\geq0\\ n+m\leq3}}\Phi^{(n,m)}z^n\bar z^m+\order{4},
 \label{SM:eq_phi_expansion}\\
 \psi&=\sum_{\substack{n,m\geq0\\ n+m\leq3}}\Psi^{(n,m)}z^n\bar z^m+\order{4},
 \label{SM:eq_psi_expansion}
\end{align}
where $\Phi^{(n,m)}$ and $\Psi^{(n,m)}$ are complex constants. Because $\phi$ and $\psi$ are real, $\Phi^{(n,m)}=\bar\Phi^{(m,n)}$ and $\Psi^{(n,m)}=\bar\Psi^{(m,n)}$.

We introduce the complex linear combination
\begin{align}
 w=q_\phi\phi+q_\psi\psi-C^{(0,0)},
 \label{SM:eq_linear_map}
\end{align}
which is related to $z$ through the near-identity transformation
\begin{align}
 w=z+h_2(z,\bar z)+h_3(z,\bar z)+\order{4},
 \label{SM:eq_near_identity}
\end{align}
where
\begin{align}
 h_2(z,\bar z)={}&C^{(2,0)}z^2+C^{(1,1)}z\bar z+C^{(0,2)}\bar z^2,
 \label{SM:eq_h2}\\
 h_3(z,\bar z)={}&C^{(3,0)}z^3+C^{(2,1)}z^2\bar z
 +C^{(1,2)}z\bar z^2+C^{(0,3)}\bar z^3.
 \label{SM:eq_h3_general}
\end{align}
The coefficients are related to those in Eqs.~\eqref{SM:eq_phi_expansion} and \eqref{SM:eq_psi_expansion} by
\begin{align}
 C^{(n,m)}=q_\phi\Phi^{(n,m)}+q_\psi\Psi^{(n,m)},
\end{align}
with
\begin{align}
 q_\phi&=\frac{\Psi^{(0,1)}}
 {\Phi^{(1,0)}\Psi^{(0,1)}-\Phi^{(0,1)}\Psi^{(1,0)}},\\
 q_\psi&=-\frac{\Phi^{(0,1)}}
 {\Phi^{(1,0)}\Psi^{(0,1)}-\Phi^{(0,1)}\Psi^{(1,0)}}.
\end{align}
As in the standard Hopf normal-form construction, the resonant cubic term is retained in the vector field rather than the coordinate transformation.
We therefore set $C^{(2,1)}=0$.

Inverting Eq.~\eqref{SM:eq_near_identity} to the third order gives
\begin{align}
 z={}&w-h_2(w,\bar w)-h_3(w,\bar w)
 +h_2(w,\bar w)\pdv{h_2}{w}(w,\bar w)
 +\bar h_2(w,\bar w)\pdv{h_2}{\bar w}(w,\bar w)
 +\order{4}.
 \label{SM:eq_inverse_transformation}
\end{align}
Hereafter, $\hat z(t)$ denotes the complex amplitude reconstructed from the measured waveforms through Eqs.~\eqref{SM:eq_linear_map} and \eqref{SM:eq_inverse_transformation}.

\subsection{Transformation coefficients from asymptotic waveforms}
\label{SM:sec1_fourier}

To reduce the number of fitted parameters, we determine most transformation coefficients from the Fourier components of the asymptotic waveforms.
We first determine the oscillation period $T$ from the interval between successive waveform peaks after relaxation and set
\begin{align}
 \omega=b-d=\frac{2\pi}{T}.
 \label{SM:eq_frequency}
\end{align}
The asymptotic waveforms are expanded as
\begin{align}
 \phi(t)&=\sum_n\phi_n e^{in\omega t},\\
 \psi(t)&=\sum_n\psi_n e^{in\omega t}.
\end{align}
Substitution of the stable limit-cycle solution $z=e^{i\omega t}$ into Eqs.~\eqref{SM:eq_phi_expansion} and \eqref{SM:eq_psi_expansion} yields
\begin{gather}
 \phi_0=\Phi^{(0,0)}+\Phi^{(1,1)},\qquad
 \psi_0=\Psi^{(0,0)}+\Psi^{(1,1)},\\
 \phi_1=\Phi^{(1,0)}+\Phi^{(2,1)},\qquad
 \psi_1=\Psi^{(1,0)}+\Psi^{(2,1)},\\
 \phi_2=\Phi^{(2,0)},\qquad \psi_2=\Psi^{(2,0)},\\
 \phi_3=\Phi^{(3,0)},\qquad \psi_3=\Psi^{(3,0)}.
\end{gather}
The condition $C^{(2,1)}=0$ gives
\begin{align}
 \Psi^{(0,1)}\Phi^{(2,1)}-\Phi^{(0,1)}\Psi^{(2,1)}=0,
\end{align}
and hence
\begin{align}
 \Phi^{(2,1)}=\bar C^{(1,2)}\Phi^{(0,1)},\qquad
 \Psi^{(2,1)}=\bar C^{(1,2)}\Psi^{(0,1)}.
\end{align}
The transformation coefficients can therefore be expressed as
\begin{align}
 q_\phi&=\frac{\psi_{-1}-C^{(1,2)}\psi_1}
 {\phi_1\psi_{-1}-\psi_1\phi_{-1}},\\
 q_\psi&=-\frac{\phi_{-1}-C^{(1,2)}\phi_1}
 {\phi_1\psi_{-1}-\psi_1\phi_{-1}},\\
 C^{(0,0)}&=q_\phi\phi_0+q_\psi\psi_0-C^{(1,1)},\\
 C^{(2,0)}&=q_\phi\phi_2+q_\psi\psi_2,\\
 C^{(0,2)}&=q_\phi\phi_{-2}+q_\psi\psi_{-2},\\
 C^{(3,0)}&=q_\phi\phi_3+q_\psi\psi_3,\\
 C^{(0,3)}&=q_\phi\phi_{-3}+q_\psi\psi_{-3}.
 \label{SM:eq_fourier_relations}
\end{align}
Thus, all transformation coefficients except $C^{(1,1)}$ and $C^{(1,2)}$ are determined from the asymptotic waveform once trial values of these two coefficients are specified.

\subsection{Inference of the intrinsic Stuart-Landau dynamics}
\label{SM:sec1_intrinsic}

After rescaling the complex amplitude so that the stable limit cycle has a unit amplitude, the intrinsic dynamics are written as
\begin{align}
 \dot z=(a+ib)z-(a+id)|z|^2z.
 \label{SM:eq_single_SL}
\end{align}
Writing $z(t)=r(t)e^{i\theta(t)}$, the analytical solution is
\begin{align}
 r(t)&=\left[1+\left(\frac{1}{r_0^2}-1\right)e^{-2at}\right]^{-1/2},
 \label{SM:eq_r_solution}\\
 \theta(t)&=(b-d)t+\frac{d}{a}\ln\frac{r(t)}{r_0}+\theta_0,
 \label{SM:eq_theta_solution}
\end{align}
where $r_0$ and $\theta_0$ specify the initial condition.

For each trial parameter set, the remaining transformation coefficients are calculated using Eq.~\eqref{SM:eq_fourier_relations}, and $\hat z(t)$ is reconstructed using Eq.~\eqref{SM:eq_inverse_transformation}.
The parameters are then inferred by minimizing the trajectory mismatch
\begin{align}
 F_{\rm cost}^{(1)}
 =\frac{1}{t_{\rm w}}
 \int_{t_{\rm s}}^{t_{\rm s}+t_{\rm w}}
 |\hat z(t)-z(t)|^2\,\dd t.
 \label{SM:eq_cost_single}
\end{align}
The fitted quantities are $C^{(1,1)}$, $C^{(1,2)}$, $a$, $b$, $r_0$, and $\theta_0$, with $b-d$ fixed by Eq.~\eqref{SM:eq_frequency}.
In the numerical implementation, the equivalent parameterization in terms of $b$ is used, with $d=b-2\pi/T$.

\subsection{Inference of coupling coefficients}
\label{SM:sec1_coupling}

For two coupled oscillators, we consider
\begin{align}
 \dot z_k={}&G_k(z_k,z_\ell)\nonumber\\
 ={}&(a+ib)z_k-(a+id)|z_k|^2z_k
 +I^{(0)}+I^{(1)}z_k+I^{(2)}\bar z_k
 +I^{(3)}z_\ell+I^{(4)}\bar z_\ell,
 \label{SM:eq_coupled_SL}
\end{align}
where $(k,\ell)=(1,2)$ or $(2,1)$.
The intrinsic parameters and coordinate transformation are fixed at the values inferred from the isolated oscillator.
The same transformation is applied to the coupled-system waveforms to obtain $\hat z_k(t)$.

The reconstructed time derivatives are evaluated using the second-order central difference
\begin{align}
 \dot{\hat z}_k(t_j)
 \simeq\frac{\hat z_k(t_{j+1})-\hat z_k(t_{j-1})}{2\Delta t}.
 \label{SM:eq_central_difference}
\end{align}
The first and last samples of each fitting interval are omitted.
No filtering or smoothing is applied. The same central-difference scheme is used in the noise test described in Sec.~\ref{SM:noisy_system}.
The accurate inference obtained under the investigated noise levels supports the practical robustness of the complete inference procedure, including the numerical differentiation step.

The real and imaginary parts of $I^{(0)},\ldots,I^{(4)}$ are inferred by minimizing the vector-field mismatch
\begin{align}
 F_{\rm cost}^{(2)}
 ={}&\sum_{\rm i.c.}\frac{1}{t_{\rm w}}
 \int_{t_{\rm s}}^{t_{\rm s}+t_{\rm w}}
 \Bigl[
 |\dot{\hat z}_1-G_1(\hat z_1,\hat z_2)|^2
 +|\dot{\hat z}_2-G_2(\hat z_2,\hat z_1)|^2
 \Bigr]\,\dd t,
 \label{SM:eq_cost_coupled}
\end{align}
where the sum is taken over waveform records generated from different initial conditions.

Equations~\eqref{SM:eq_cost_single} and \eqref{SM:eq_cost_coupled} respectively quantify the trajectory mismatch in the isolated oscillator and the vector-field mismatch in the coupled system. 
They provide a simple operational implementation of the normal-form-based inference framework. 
The central coordinate-reconstruction principle does not depend on these particular objective functions or on the numerical optimizer used to minimize them.

\subsection{Numerical optimization}
\label{SM:sec1_optimization}

Both objective functions are minimized using the \texttt{differential\_evolution} routine in SciPy~\cite{scipy2019}. 
Unless otherwise specified, we use
\begin{align}
 \texttt{tol}=10^{-10},\qquad
 \texttt{maxiter}=1000,\qquad
 \texttt{seed}=0,
\end{align}
with the remaining options set to the SciPy defaults. 
The system-specific parameter bounds are reported in Sec.~\ref{SM:sec_vdp_settings} for the VDP system and Sec.~\ref{SM:sec_fluid_settings} for the collapsible-channel system.

\subsection{Data-window stability analysis}
\label{SM:sec1_window}

The start time $t_{\rm s}$ and duration $t_{\rm w}$ specify the data segment used for inference. 
Rather than selecting a unique optimum solely from the minimum value of the fitting objective, we evaluate the sensitivity of the inferred parameters over the $(t_{\rm s},t_{\rm w})$ plane using
\begin{align}
 F_{\rm cost}^{(\rm hyp)}
 =\left\|\frac{\partial\bm c}{\partial t_{\rm s}}\right\|^2
 +\left\|\frac{\partial\bm c}{\partial t_{\rm w}}\right\|^2,
 \label{SM:eq_cost_hyper}
\end{align}
where the derivatives are evaluated by forward differences.
For the isolated-oscillator inference,
\begin{align}
 \bm c=\left(C^{(1,1)},C^{(1,2)},a,d\right),
\end{align}
and for the coupled-oscillator inference,
\begin{align}
 \bm c=\left(I^{(0)},I^{(1)},I^{(2)},I^{(3)},I^{(4)}\right).
\end{align}

The diagnostic $F_{\rm cost}^{(\rm hyp)}$ is not intended to define a universal or uniquely optimal pair of $(t_{\rm s},t_{\rm w})$.
Instead, it identifies broad regions in which the inferred parameters are insensitive to moderate variations in the data window.
Selecting a broad low-sensitivity region is preferable to selecting an isolated point with a small value of $F_{\rm cost}^{(\rm hyp)}$, because the latter may reflect the particular data segment rather than a robustly identifiable parameter set.

For the isolated-oscillator inference, the stable regions found here correspond to windows containing a substantial part of the amplitude-relaxation process, which provides information unavailable from the asymptotic limit cycle alone.
For the coupled-oscillator inference, the diagnostic similarly identifies windows containing sufficient transient dynamics for stable estimation of the coupling coefficients.

The system-specific time intervals, sampling intervals, and resulting stability maps are given in Sec.~\ref{SM:sec_system_settings}.

\section{System-specific simulation and inference settings}
\label{SM:sec_system_settings}

This section summarizes the numerical simulations and the system-dependent inference settings used for the coupled van der Pol (VDP) and collapsible-channel systems. 
The common coordinate-reconstruction procedure, objective functions, numerical differentiation scheme, optimization method, and data-window stability diagnostic are described in Sec.~\ref{SM:Inference_detail}.

\subsection{Coupled van der Pol oscillators}
\label{SM:sec_vdp_settings}

We numerically solved the coupled VDP equations
\begin{align}
 \ddot{x}_k={}&(\mu-x_k^2)\dot{x}_k-x_k
 -I_x(x_k-x_\ell)-I_v(\dot{x}_k-\dot{x}_\ell),
 \label{SM:eq_vdp_setup}
\end{align}
where $(k,\ell)=(1,2)$ or $(2,1)$.

To prepare the initial conditions, we first integrated the uncoupled system with $I_x=I_v=0$ until $t=0$, allowing each oscillator to approach its limit cycle. 
We set $(x_1,\dot{x}_1)=(0.1,0)$ at $t=-300$ and $(x_2,\dot{x}_2)=(0.1,0)$ at $t=-300+\tau^{(0)}$, where $\tau^{(0)}$ denotes the initial time delay between the oscillators. 
At $t=0$, the coupling parameters were switched from zero to their target values, and the simulations were continued until $t=300$.

Unless otherwise specified, we set $\mu=0.2$ and varied the coupling parameters over
\begin{align}
 0\leq I_x\leq0.2,
 \qquad
 0\leq I_v\leq0.1.
\end{align}
The initial delay $\tau^{(0)}$ was sampled at 21 equally spaced points over one intrinsic period $T$.

For the isolated-oscillator inference, the Fourier coefficients of the asymptotic waveform were computed from $-T\leq t\leq0$ with a sampling interval of $0.01$. 
The transient data used to infer the remaining transformation coefficients and intrinsic SL parameters were sampled from $t=-300$ to $-295$ with a sampling interval of $0.1$. 
Thus, the selected data window was
\begin{align}
 t_{\rm s}=-300,
 \qquad
 t_{\rm w}=5.
\end{align}

For the coupled-system inference, waveform data corresponding to the different initial delays were sampled from $t=17.5$ to $35.5$ with a sampling interval of $0.1$, corresponding to
\begin{align}
 t_{\rm s}=17.5,
 \qquad
 t_{\rm w}=18.
\end{align}

The parameter bounds used for the isolated-oscillator inference were
\begin{align}
 &\Re C^{(1,1)},\Im C^{(1,1)},
 \Re C^{(1,2)},\Im C^{(1,2)}\in[-0.5,0.5],\\
 &a\in[0.01,1],\\
 &b\in[2\pi/T-0.1,\,2\pi/T+0.1],\\
 &r_0\in[0.01,1],\\
 &\theta_0\in[0,2\pi].
\end{align}
For the coupled-system inference, the real and imaginary parts of $I^{(0)},\ldots,I^{(4)}$ were searched over $[-2,2]$.

\begin{figure}[t]
 \includegraphics[width=10cm]{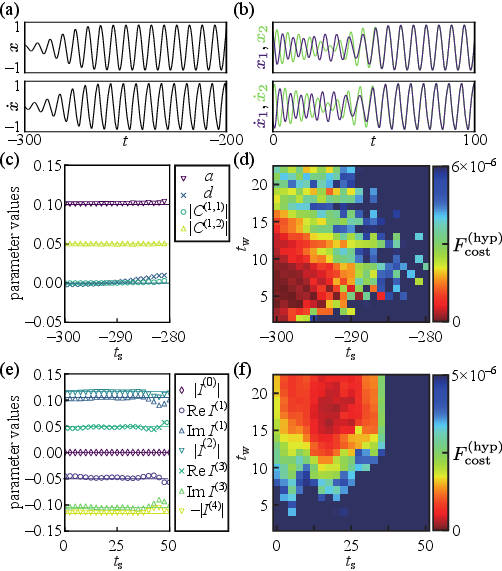}
 \caption{
 Data-window dependence of the inference for the VDP system.
 (a) Typical time series of $x$ and $\dot{x}$ used to infer the parameters of a single oscillator.
 (b) Typical time series of $x_k$ and $\dot{x}_k$ ($k = 1, 2$) used to infer the parameters of coupled oscillators.
 (c) Inferred intrinsic parameters as functions of $t_{\rm s}$ at fixed $t_{\rm w}$.
 (d) Stability diagnostic $F_{\rm cost}^{(\rm hyp)}$ for the isolated-oscillator inference in the $(t_{\rm s},t_{\rm w})$ plane.
 (e) Inferred coupling coefficients as functions of $t_{\rm s}$ at fixed $t_{\rm w}$.
 (f) Stability diagnostic for the coupled-system inference.
 Regions with smaller $F_{\rm cost}^{(\rm hyp)}$ indicate weaker sensitivity of the inferred parameters to the data window.
 }
 \label{fig:S1}
\end{figure}

Figures~\ref{fig:S1}(a) and \ref{fig:S1}(b) show the typical time series used to infer the parameters.
Figures~\ref{fig:S1}(c) and \ref{fig:S1}(e) show the dependence of the inferred parameters on the data window. 
Broad regions with small values of the stability diagnostic $F_{\rm cost}^{(\rm hyp)}$ are observed for both inference stages [Figs.~\ref{fig:S1}(d) and \ref{fig:S1}(f)]. 
The windows specified above were selected from these stable regions. 
For the isolated oscillator, the appropriate window length is comparable to the relaxation time, indicating that the transient amplitude dynamics provide essential information for parameter inference.

\subsection{Coupled collapsible channels}
\label{SM:sec_fluid_settings}

We applied the inference procedure to oscillatory flows in two collapsible channels that merge into a single outlet channel. 
The governing equations and numerical method for the hydrodynamic simulations are described in Ref.~\cite{Araya2024}. 
Here, we summarize the conditions used to generate the waveform data and perform the inference.

The stretching stiffness of the elastic walls was set to
\begin{align}
 K_s=1200,
\end{align}
placing the isolated channel slightly above the supercritical Hopf bifurcation. 
The channel-coupling parameter $\delta$ was varied from $0$ to $10$ in increments of $0.2$. 
At $\delta=0$, the two channels are independent.

To prepare initial conditions with different phase offsets, the upper elastic wall was initialized to be parallel to the $x_1$ axis at $t=-700$, whereas the lower wall was initialized in the same configuration at $t=-700+\tau^{(0)}$. 
The initial delay $\tau^{(0)}$ was sampled at 34 equally spaced points over one period $T$ of an isolated oscillatory flow. 
The system was simulated with $\delta=0$ until $t=0$. At $t=0$, $\delta$ was switched to its target value, and the simulation was continued until $t=750$.

For each oscillator, the coordinates $(X,Y)$ of the lowest point of the elastic wall were used as the two waveform observables. 
The Fourier coefficients of the asymptotic isolated-channel waveform were computed from $-T\leq t\leq0$ with a sampling interval of $0.0025$. 
The remaining transformation coefficients and intrinsic SL parameters were inferred from the transient data sampled from $t=-675$ to $-625$ with a sampling interval of $0.25$, corresponding to
\begin{align}
 t_{\rm s}=-675,
 \qquad
 t_{\rm w}=50.
\end{align}

For the coupled-channel inference, waveform data corresponding to the different initial phase offsets were sampled from $t=80$ to $180$ with a sampling interval of $0.25$. The corresponding data window was
\begin{align}
 t_{\rm s}=80,
 \qquad
 t_{\rm w}=100.
\end{align}

The parameter bounds for the isolated-channel inference were
\begin{align}
 &\Re C^{(1,1)},\Im C^{(1,1)},
 \Re C^{(1,2)},\Im C^{(1,2)}\in[-0.5,0.5],\\
 &a\in[0.001,0.1],\\
 &b\in[1,3],\\
 &r_0\in[0.01,1],\\
 &\theta_0\in[0,2\pi].
\end{align}
For the coupled-channel inference, the real and imaginary parts of $I^{(0)},\ldots,I^{(4)}$ were searched over $[-2,2]$.

\begin{figure}[t]
 \includegraphics[width=10cm]{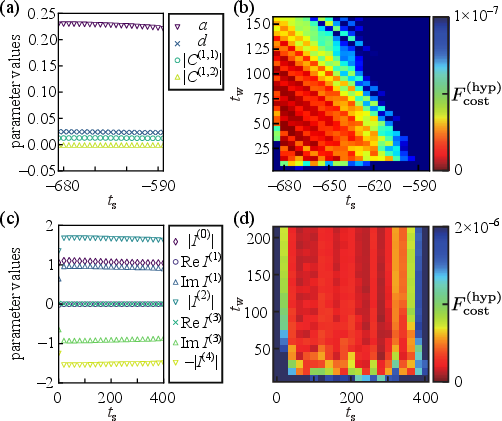}
 \caption{
 Data-window dependence of the inference for the collapsible-channel system.
 (a) Inferred intrinsic parameters as functions of $t_{\rm s}$ at fixed $t_{\rm w}$.
 (b) Stability diagnostic $F_{\rm cost}^{(\rm hyp)}$ for the isolated-channel inference in the $(t_{\rm s},t_{\rm w})$ plane.
 (c) Inferred coupling coefficients as functions of $t_{\rm s}$ at fixed $t_{\rm w}$.
 (d) Stability diagnostic for the coupled-channel inference.
 Regions with smaller $F_{\rm cost}^{(\rm hyp)}$ indicate weaker sensitivity of the inferred parameters to the data window.
 }
 \label{fig:S1+}
\end{figure}

Figures~\ref{fig:S1+}(a) and \ref{fig:S1+}(c) show the dependence of the inferred parameters on the selected data window. 
The corresponding stability maps [Figs.~\ref{fig:S1+}(b) and \ref{fig:S1+}(d)] contain broad regions with small $F_{\rm cost}^{(\rm hyp)}$. 
The inference windows specified above were selected from these stable regions.

\section{Stability analysis for coupled van der Pol oscillators \label{SM:VDP_analysis}}
We analytically derive the Stuart-Landau equation from the van der Pol (VDP) equation using a near-identity transformation. By introducing $v\equiv\dot x$, the VDP equation is rewritten as a two-dimensional dynamical system,
\begin{align}
\dot x &=v, \\
\dot v &=(\mu-x^2)v-x.
\end{align}
One of the eigenvalues $\lambda$ at  the fixed point $(x,v)=(0,0)$ is obtained as 
\begin{align}
\lambda=\frac{\mu}{2}+i\sqrt{1-\frac{\mu^2}{4}}
\end{align}
by the linear stability analysis. 
Since we consider the onset of the Hopf bifurcation point, we assume $1 - \mu^2/4 > 0$. 
A right eigenvector $\bm p= {}^t(p_x,p_v)$ corresponding to $\lambda$ satisfies $p_v=\lambda p_x$. 
Defining the left eigenvector $\bm{q}=(q_x,q_v)$ so that 
\begin{align}
p_xq_x+p_vq_v=1 \label{normal}
\end{align}
and 
\begin{align}
\bar{p}_xq_x+\bar{p}_vq_v=0,
\end{align}
we obtain 
\begin{align}
\mqty(q_x, q_v)=\frac{\bar p_x}{|p_x|^2(\bar\lambda-\lambda)}\mqty(\bar\lambda,-1).
\end{align}
It should be noted that the $x$-component of the right eigenvector $p_x$ can be arbitrarily chosen. 
Considering that $p_x$ is a complex number, the absolute value of $p_x$ determines the ratio of $\left|\bm{p}\right|$ and $\left|\bm{q}\right|$, and the argument of $p_x$ determines the correspondence of the phase origins of the VDP equation and the Stuart-Landau equation.

We define the complex amplitude $z$ by $z\equiv q_xx+q_vv$ using the right eigenvector. 
Then, $x$ and $v$ are expressed as $x=zp_x+\bar z\bar p_x$, and $v=zp_v+\bar z\bar p_v$. 
The time-evolution equation for $z$ is thus
\begin{align}
    \dot{z}
    ={}&\lambda z-\lambda q_vp_x^3z^3-q_vp_x^2\bar p_x(\bar\lambda+2\lambda)z^2\bar z -q_vp_x\bar p_x^2(2\bar \lambda+\lambda)z\bar z^2-\bar\lambda q_v\bar p_x^3\bar z^3.
\end{align}
Applying the near-identity transformation
\begin{align}
z\mapsto z+C^{(3,0)}z^3+C^{(1,2)}z\bar z^2+C^{(0,3)}\bar z^3,
\end{align}
we obtain
\begin{align}
    &C^{(3,0)}=\frac{i\mu}{4\sqrt{1-\mu^2/4}}\frac{p_x^2}{|p_x|^2},\\
    &C^{(1,2)}=\left(-\frac{\mu^2}{4}+i\frac{\mu+\mu^3/2}{4\sqrt{1-\mu^2/4}}\right)\frac{\bar p_x^2}{|p_x|^2},\\
    &C^{(0,3)}=\left(-\frac{\mu^2}{32-6\mu^2}+i\frac{\mu-\mu^3/8}{(8-3\mu^2/2)\sqrt{1-\mu^2/4}}\right)\frac{\bar p_x^4}{|p_x|^4}.
\end{align}
Under this transformation, time evolution of $z$ is given as
\begin{align}
    \dot{z}={}&\left(\frac{\mu}{2}+i\sqrt{1-\frac{\mu^2}{4}}\right)z-\left(\frac12-i\frac{3\mu}{4\sqrt{1-\mu^2/4}}\right)|p_x|^2z^2\bar z+\order{5}.
\end{align}

Choosing $|p_x|^2=\mu$ so that the modulus of the complex amplitude is $1$ on the limit cycle solution, we get
\begin{align}
    \dot{z}={}&\left(\frac{\mu}{2}+i\right)z-\left(\frac{\mu}{2} - i \frac{3\mu^2}{4\sqrt{1 - \mu^2/4}} \right) z^2\bar z+\order{5}.
\end{align}
The transformation between $z$ and $(x,v)$ has still one free parameter that reflects the phase relationship. 
If we set it so that $\arg z = 0$ at $v=0$ with $x > 0$, then the explicit transformation is given by setting $p_x = \sqrt{\mu(1 -\mu^2/4)}+ i \mu \sqrt{\mu}/2 $.
In this case, the linear transformation is given as
\begin{align}
z = \frac{x - (\mu/2) v}{2\sqrt{\mu(1-\mu^2/4)}} - i \frac{v}{2\sqrt{\mu}}, \label{trans1}
\end{align}
and
\begin{align}
\left(x,v \right) = \left(\sqrt{\mu(4-\mu^2)}\Re z - \mu \sqrt{\mu} \Im z, -2 \sqrt{\mu} \Im z\right). \label{trans2}
\end{align}
The coefficients for the near-identity transformation are also explicitly given as
\begin{align}
    &C^{(3,0)}= -\frac{\mu^2}{4} + i \frac{\mu(2 - \mu^2)}{4 \sqrt{4 - \mu^2}},\\
    &C^{(1,2)}=\frac{\mu^4}{4} + i \frac{2 + 2\mu^2 - \mu^4}{4 \sqrt{4 - \mu^2}},\\
    &C^{(0,3)}= \frac{\mu^2(7 - 3\mu^2)}{32 - 6\mu^2} + i \frac{\mu(8 - 13 \mu^2 + 3 \mu^4)}{2(16 - 3\mu^2)\sqrt{4-\mu^2}}. 
\end{align}
Considering that our inference method uses the third-order near-indentity transformation, $\mu$ is a small parameter and the analysis is precise upto the order of $\mu^1$. 
Thus, for the comparison to the inference results, we adopt
$a = \mu/2$, $b = 1$, and $d = 0$ by neglecting the terms of $\mathcal{O}(\mu^2)$.

Next, we transform the coupled VDP equations with setting $\dot x_1=v_1,\ \dot x_2=v_2$. 
Applying the near-identity transformation defined by the single oscillator, $(x_1,v_1)$ and $(x_2,v_2)$ are mapped to complex amplitudes $z_1$ and $z_2$, respectively, yielding
\begin{align}
    &\dot z_1=\left(\frac{\mu}{2}+i\right)z_1-\frac{\mu}{2}z_1^2\bar z_1+I(z_1,z_2)+\order{\mu^2},\\
    &\dot z_2=\left(\frac{\mu}{2}+i\right)z_2-\frac{\mu}{2}z_2^2\bar z_2+I(z_2,z_1)+\order{\mu^2},
\end{align}
where the linear part of the interaction term $I$ is
\begin{align}
    I(z_1,z_2)={}&\left(\frac{I_v}{2}-i\frac{2I_x+\mu I_v}{4}\right)(z_2-z_1)-\left(\frac{I_v}{2}+i\frac{2I_x+\mu I_v}{4}\right)\frac{\bar p_x^2}{|p_x|^2}(\bar z_2-\bar z_1).
\end{align}

Thus, $I^\qty(0) = 0$, $I^\qty(1) = -I_v/2+i (2I_x+\mu I_v)/4$, and $I^\qty(3) = I_v/2 -i (2I_x+\mu I_v)/4$. 
The coefficients $I^\qty(2)$ and $I^\qty(4)$ for the complex conjugates $\bar{z}_1$ and $\bar{z}_2$ depend on $p_x$, but they have relation $I^\qty(2) = -I^\qty(4)$ and $\left| I^\qty(1)\right| = \left| I^\qty(2)\right| = \left|I^\qty(3)\right| = \left| I^\qty(4)\right|$ independent of the choice of $p_x$. 
In the case of the transformation in Eqs.~\eqref{trans1} and \eqref{trans2}, we obtain $I^\qty(2) = -I^\qty(4) = (\mu I_x + I_v)/2 + i (2I_x - \mu I_v)/4$.

\section{Detailed calculation for the linear stability analyses of coupled Stuart-Landau equations \label{SM:cSL_stability}}

Here, we analyze the stability of the coupled Stuart-Landau equation (Eq.~(\ref{coupledSL}) in the main text) under the assumption that the non-resonance terms $I^\qty(0)$, $I^\qty(2)\bar z_1$ and $I^\qty(4)\bar z_2$ are neglected.
It should be noted that the limit cycles corresponding to the in-phase and anti-phase states in the phase space of $(z_1, z_2)$ correspond to fixed points in the $(r_+, r_-, \theta_-)$ space. 
First, we obtain the forms with polar coordinates $z_1=r_1 e^{i\theta_1}$ and $z_2=r_2e^{i\theta_2}$ expressed as
\begin{align}
  \dot r_+={}&a(r_+-r_+^3-3r_+r_-^2)+\Re I^\qty(1)r_+ +\Re I^\qty(3)r_ +\cos\theta_- -\Im I^\qty(3)r_-\sin\theta_-,
\end{align}
\begin{align}
  \dot r_-={}&a(r_--3r_+^2r_--r_-^3)+\Re I^\qty(1)r_- -\Re I^\qty(3)r_-\cos\theta_-+\Im I^\qty(3)r_+\sin\theta_-,
\end{align}
\begin{align}
  \dot\theta_-={}& - 4 d r_+ r_- - \Re I^\qty(3)\frac{2r_+^2+2r_-^2}{r_+^2-r_-^2}\sin\theta_- -\Im I^\qty(3)\frac{4r_+r_-}{r_+^2-r_-^2}\cos\theta_-,
\end{align}
where $r_\pm=(r_1\pm r_2)/2$ and $\theta_-=\theta_1-\theta_2$. We apply the linear stability analysis to the limit cycle, where $\dot r_+=0$, $\dot r_-=0$, and $\dot \theta_-=0$. 
The solution corresponding to the in-phase synchronization is given as 
$(r_+^*,r_-^*,\theta_-^*)=(\sqrt{1+(\mathrm{Re}I^\qty(1)+\mathrm{Re}I^\qty(3))/a},0,0)$, and that corresponding to the anti-phase synchronization is given as 
$(r_+^*,r_-^*,\theta_-^*)=(\sqrt{1 +(\mathrm{Re}I^\qty(1) - \mathrm{Re}I^\qty(3))/a},0,\pi)$. The eigenvalues of the Jacobi matrix for the linearized equations are obtained as
\begin{align}
    \lambda^\mathrm{(amp)}_\mathrm{IP} =& -2 \left(a + \mathrm{Re}I^\qty(1) +\mathrm{Re}I^\qty(3)\right), \nonumber \\
    & -a - \mathrm{Re}I^\qty(1)-3\mathrm{Re}I^\qty(3) \pm \sqrt{\left( a + \mathrm{Re}I^\qty(1)+\mathrm{Re}I^\qty(3)\right)^2  - 4\mathrm{Im}I^\qty(3)\left\{\mathrm{Im}I^\qty(3) + d \left(1 + \frac{\mathrm{Re}I^\qty(1) + \mathrm{Re}I^\qty(3)}{a}\right)\right\}} \label{ev}
\end{align}
for the in-phase synchronization, and
\begin{align}
    \lambda^\mathrm{(amp)}_\mathrm{AP} =& -2\left( a + \mathrm{Re}I^\qty(1) - \mathrm{Re}I^\qty(3)\right), \nonumber \\
    & -a - \mathrm{Re}I^\qty(1) + 3\mathrm{Re}I^\qty(3) \pm \sqrt{\left(a + \mathrm{Re}I^\qty(1) - \mathrm{Re}I^\qty(3)\right)^2 - 4\mathrm{Im}I^\qty(3)\left\{\mathrm{Im}I^\qty(3) - d \left(1 + \frac{\mathrm{Re}I^\qty(1) - \mathrm{Re}I^\qty(3)}{a}\right)\right\}} \label{evdirect}
\end{align}
for the anti-phase synchronization. 
It should be noted that the Floquet exponents for the original system are given by these eigenvalues and 0, where 0 corresponds to the neutral mode in the direction of the trajectory.

From the results shown in Sec.~\ref{SM:VDP_analysis}, the parameters $a$, $c$, $d$, $I^\qty(1)$, and $I^\qty(3)$ are given for the coupled VDP equation.
Under the assumption that the non-resonant terms $I^\qty(0)$, $I^\qty(2)z$, $I^\qty(4)\bar{z}$ in the coupling terms, are neglected, the eigenvalues for the in-phase and anti-phase synchronizations are expressed as
\begin{align}
    \lambda^\mathrm{(amp)}_\mathrm{IP} =& - \mu, \, -\frac{\mu}{2} - I_v\pm\sqrt{\left(\frac{\mu}{2} \right)^2-\left(I_x+ \frac{\mu I_v}{2} \right)^2},
\end{align}
and
\begin{align}
    \lambda^\mathrm{(amp)}_\mathrm{AP} =& - \mu + 2 I_v, \, -\frac{\mu}{2}+2I_v\pm \sqrt{\left(\frac{\mu}{2}-I_v \right)^2-\left(I_x+ \frac{\mu I_v}{2}\right)^2}, \label{lam_anti_amp}
\end{align}
respectively.
For the parameter region $\mu > 0$, $I_x > 0$, and $I_v > 0$, which corresponds to the region shown in Fig.~\ref{fig:3} in the main text, all of the real parts of the eigenvalues for the in-phase synchronization are always negative, indicating that the in-phase synchronization is stable.
In contrast, the anti-phase synchronization mode becomes unstable via a Neimark-Sacker bifurcation at
\begin{align}
    I_v=\frac{\mu}{4}, \label{NS}
\end{align}
and a pitchfork bifurcation at
\begin{align}
    \qty(I_x+\frac{\mu I_v}{2})^2+3\qty(I_v-\frac{\mu}{6})^2=\frac{\mu^2}{12}. \label{full_PF}
\end{align}

Next, we consider the adiabatic approximation with respect to the amplitudes. 
We assume $\dot{r}_\pm = 0$ and obtain a time evolution equation for $\theta_-$, which is valid close to the fixed points corresponding to the in-phase and anti-phase states ($\theta_- = 0$ and $\pi$, respectively). 
The linearized equation around the fixed point $\theta_- = 0$ is given as
\begin{align}
    \dot{\delta \theta_-} =& \left[ -4d r_{+0} K_0 -  4 \mathrm{Im}I^\qty(3) \frac{K_0}{r_{+0}} - 2 \mathrm{Re}I^\qty(3) \right] \delta \theta_- ,
\end{align}
where $r_{+0}$ is the steady-state solution $r_{+0} = \sqrt{1 + (\mathrm{Re}I^\qty(1) + \mathrm{Re}I^\qty(3))/a}$ and $K_0$ is given as
\begin{align}
K_0 =\frac{ r_{+0}  \mathrm{Im}I^\qty(3)}{2 (a + \mathrm{Re}I^\qty(1) + 2\mathrm{Re}I^\qty(3))}.
\end{align}
Thus, the eigenvalue for the linearized equation is given as
\begin{align}
    \lambda^\mathrm{(ad)}_\mathrm{IP} =& - \frac{2 \mathrm{Im}I^\qty(3) (a d + d \mathrm{Re}I^\qty(1) + d \mathrm{Re}I^\qty(3) + a  \mathrm{Im}I^\qty(3))}{a( a + \mathrm{Re}I^\qty(1) + 2\mathrm{Re}I^\qty(3))}  - 2 \mathrm{Re}I^\qty(3) \label{evinad}
\end{align}
The linearized equation around the fixed point $\theta_- = \pi$, i.e., anti-phase synchronization, is also given in the same manner as
\begin{align}
    \dot{\delta \theta_-} =& \left[ -4d r_{+\pi} K_\pi +  4 \mathrm{Im}I^\qty(3) \frac{K_\pi}{r_{+\pi}} + 2 \mathrm{Re}I^\qty(3) \right] \delta \theta_- ,
\end{align}
where $r_{+\pi}$ is the steady-state solution $r_{+\pi} = \sqrt{1 + (\mathrm{Re}I^\qty(1) - \mathrm{Re}I^\qty(3))/a}$ and $K_\pi$ is given as
\begin{align}
K_\pi =-\frac{ r_{+\pi}  \mathrm{Im}I^\qty(3)}{2 (a + \mathrm{Re}I^\qty(1) - 2\mathrm{Re}I^\qty(3))}.
\end{align}
Thus, the eigenvalue for the linearized equation is given as
\begin{align}
    \lambda^\mathrm{(ad)}_\mathrm{AP} =& \frac{2 \mathrm{Im}I^\qty(3) (a d + d \mathrm{Re}I^\qty(1) - d \mathrm{Re}I^\qty(3) - a  \mathrm{Im}I^\qty(3))}{a( a + \mathrm{Re}I^\qty(1) - 2\mathrm{Re}I^\qty(3))} + 2 \mathrm{Re}I^\qty(3). \label{evantiad} 
\end{align}
For the coupled VDP equations, the eigenvalues for the in-phase and anti-phase states are calculated as
\begin{align}
 \lambda^\mathrm{(ad)}_\mathrm{IP} = - I_v - \frac{(2I_x + \mu I_v)^2}{4(\mu + I_v)}, 
\end{align}
and
\begin{align}
 \lambda^\mathrm{(ad)}_\mathrm{AP} = I_v - \frac{(2I_x + \mu I_v)^2}{4(\mu - 3I_v)}, \label{lam_anti_ad}
\end{align}
respectively. 
Thus, for $\mu>0$, $I_x >0$, and  $I_v > 0$, the in-phase synchronization is always stable, while the anti-phase synchronization becomes unstable at the condition given by Eq.~\eqref{full_PF} via a pitchfork bifurcation.

\begin{figure}[tb]
  \includegraphics{"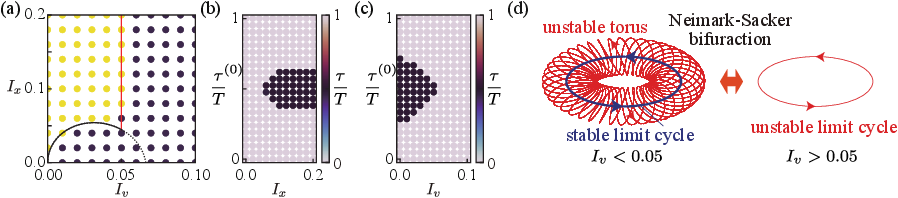"}
  \caption{
    Numerical results for the coupled van der Pol (VDP) equations.
    (a)~Phase diagram for the anti-phase synchrony, where the yellow and purple circles indicate stable and unstable anti-phase states, respectively.
    From the analysis using the coupled Stuart-Landau equations with non-resonance terms neglected, the anti-phase synchrony is stable when all the real values of $\lambda_\mathrm{AP}^{(\mathrm{amp})}$ in Eq.~\eqref{lam_anti_amp} are negative.
    The stability boundary is shown by the black line [Eq.~\eqref{full_PF}, the pitchfork bifurcation] and red solid line [Eq.~\eqref{NS}, the Neimark-Sacker bifurcation]. 
    In contrast, using the equation with the adiabatic approximation, the stability boundary is given by $\lambda_\mathrm{AP}^\mathrm{(ad)} = 0$; i.e. Eq.~\eqref{full_PF}, which is represented by the combination of the black solid and dashed lines.
    (b, c)~Normalized phase difference $\tau/T$ in the final state depending on (ii)~$\tau^\qty(0)/T$ and $I_x$ at $I_v=0.03$ and (iii)~$\tau^\qty(0)/T$ and $I_v$ at $I_x=0.16$.
    $\tau^\qty(0)/T$ is the initial normalized phase difference between two oscillators and $T$ is the intrinsic period of the single oscillator.
    (d) Schematic illustration of the Neimark-Sacker bifurcation. The stable limit cycle collides with an unstable torus and destabilizes at the bifurcation point.
  \label{fig:4}}
\end{figure}

Finally, we consider the case using the coupled phase equations. 
From a single Stuart-Landau equation, we obtain the phase equation as
\begin{align}
    \dot{\theta} = b - d, 
\end{align}
since the stable amplitude is $1$. 
Then, we obtain the coupled equation as
\begin{align}
    \dot{\theta}_k =& b - d + \mathrm{Im}I^\qty(1) + \mathrm{Re}I^\qty(3) \sin(\theta_\ell - \theta_k) + \mathrm{Im}^\qty(3) \cos(\theta_\ell - \theta_k).
\end{align}
By setting $\theta_- = \theta_1 - \theta_2$, we obtain
\begin{align}
    \dot{\theta}_- = -2 \mathrm{Re}I^\qty(3) \sin \theta_-.
\end{align}
The eigenvalues for in-phase and anti-phase synchronizations are given as
\begin{align}
    \lambda^\mathrm{(phase)}_\mathrm{IP} = - 2 \mathrm{Re}I^\qty(3),
\end{align}
and
\begin{align}
    \lambda^\mathrm{(phase)}_\mathrm{AP} = 2 \mathrm{Re}I^\qty(3),
\end{align}
respectively. 
For the coupled VDP equations, $\mathrm{Re}I^\qty(3) = I_v/2$, and therefore in-phase synchronization is stable and anti-phase synchronization is unstable for $\mu>0$, $I_x > 0$, and $I_v > 0$. 
The analysis using the phase equation is only correct when the interaction is sufficiently weak; it cannot predict the stable anti-phase synchrony.

Figure~\ref{fig:4} shows the stability boundaries of the anti-phase synchronization obtained from the eigenvalues without approximation and those with the adiabatic approximation.
The former can anticipate both the Neimark-Sacker and pitchfork bifurcations, while the latter can anticipate only the pitchfork bifurcation.
The boundary tangentially touches $I_x$-axis at the origin, which shows that the anti-phase synchronization is unstable for the small $I_x$ and $I_v$. 

As shown in Fig.~\ref{fig:4}(a),
the bifurcation line obtained from the linear stability analysis is quantitatively consistent with the numerical simulation results.
We also show the bifurcation line obtained from the adiabatic approximation with respect to the amplitudes using black solid and dashed lines.
We can reproduce the pitchfork bifurcation line using the adiabatic approximation, while we fail to lead the Neimark-Sacker bifurcation line.

\section{Validity ranges with respect to coupling strength and distance from the Hopf bifurcation}
\label{SM:sec_validity}

To quantify the validity ranges of the inferred reduced model, we examined its dependence on the coupling strengths $I_x$ and $I_v$ and on the bifurcation parameter $\mu$.
For each parameter set, we compared the inferred coefficients and oscillation frequencies with their leading-order analytical values and compared the Floquet exponents of the inferred coupled Stuart-Landau (SL) equations with those computed directly from the original coupled van der Pol (VDP) equations.

\begin{figure}[tb]
  \includegraphics{"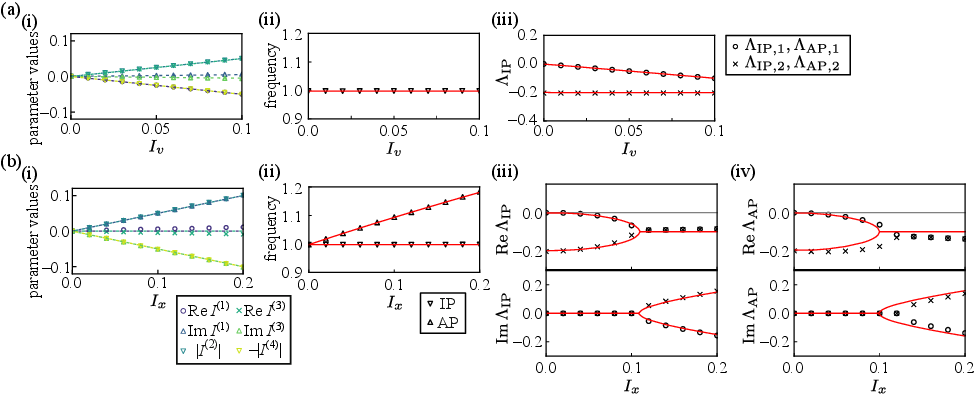"}
  \caption{
  Validity of the inference under variations in the coupling strength in the coupled van der Pol (VDP) equations.  
  (a) Scan of $I_v$ at $I_x=0$ and fixed $\mu=0.2$.
  (b) Scan of $I_x$ at $I_v=0$ and fixed $\mu=0.2$.
  (i) Inferred parameters (markers) and their leading-order analytical values (dashed lines); the colors of the dashed lines correspond to those of the markers.
  (ii) Oscillation frequencies of the in-phase (IP) and anti-phase (AP) synchronized solutions.
  (iii) Floquet exponents of the IP solution.
  (iv) Floquet exponents of the AP solution.
  In panels (ii)--(iv), markers denote values obtained from the inferred coupled SL equations, whereas solid lines denote values computed from the original coupled VDP equations.
  Both the real and imaginary parts of the Floquet exponents are shown.
  }
  \label{fig:E}
\end{figure}

\begin{figure}[tb]
  \includegraphics{"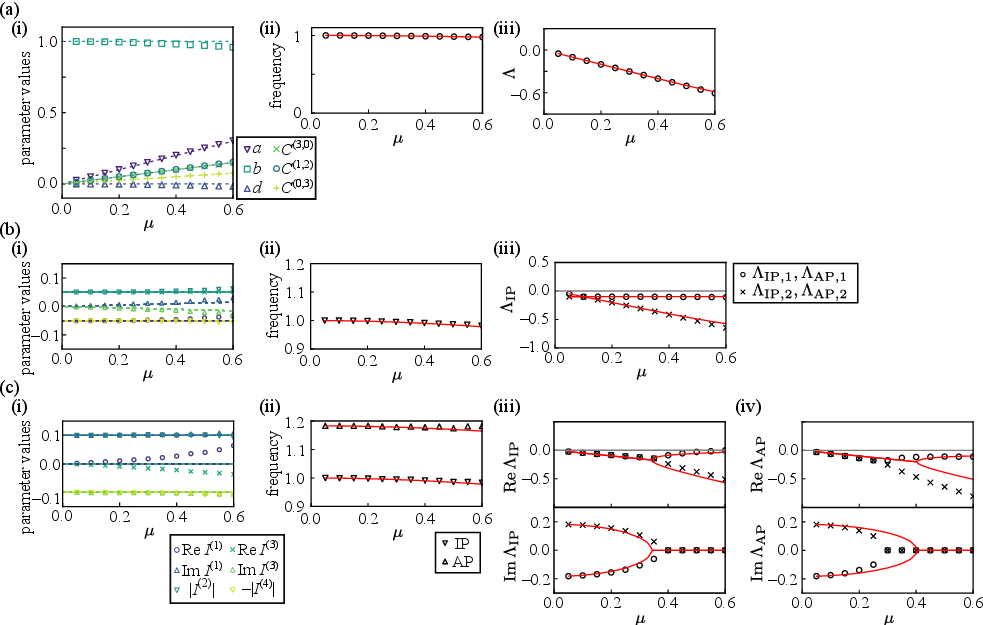"}
  \caption{
  Validity of the inference as the system moves away from the Hopf bifurcation in the coupled van der Pol (VDP) equations.  
  The bifurcation parameter $\mu$ is varied for
  (a) an isolated VDP oscillator,
  (b) coupled VDP oscillators with $I_x=0$ and $I_v=0.1$, and
  (c) coupled VDP oscillators with $I_x=0.2$ and $I_v=0$.
  (i) Inferred parameters (markers) and their leading-order analytical values (dashed lines); the colors of the dashed lines correspond to those of the markers.
  (ii) Oscillation frequencies, where applicable.
  (iii) Floquet exponents of the IP solution.
  (iv) Floquet exponents of the AP solution.
  For the coupled systems, markers in panels (ii)--(iv) denote values obtained from the inferred coupled SL equations, whereas solid lines denote values computed from the original coupled VDP equations.
  Both the real and imaginary parts of the Floquet exponents are shown.
  }
  \label{fig:F}
\end{figure}

\subsection{Periodic solutions of the coupled VDP equations}

We performed a numerical stability analysis of periodic solutions of the coupled VDP equations
\begin{align}
    \dot x_k &= v_k,\\
    \dot v_k &= (\mu-x_k^2)v_k-x_k
    -I_x(x_k-x_\ell)-I_v(v_k-v_\ell),
    \label{SM:eq_vdp_stability}
\end{align}
where $(k,\ell)=(1,2)$ or $(2,1)$.
For convenience, we introduce the complex variable
\begin{align}
    z_k=x_k-i v_k,
\end{align}
so that Eq.~\eqref{SM:eq_vdp_stability} can be expressed as
\begin{align}
    \dot z_k=G_k'(z_k,z_\ell).
\end{align}

The periodic solutions were represented by truncated Fourier series.
The IP synchronized solution was written as
\begin{align}
    z_1^*(t)=z_2^*(t)
    =\sum_{n=-N}^{N}c_n\exp(in\omega t).
    \label{SM:eq_IP_fourier}
\end{align}
Because the diffusive coupling terms vanish when $x_1=x_2$ and $v_1=v_2$, the IP synchronized solution coincides with the limit-cycle solution of an isolated VDP oscillator.
The AP synchronized solution was written as
\begin{align}
    z_1^*(t)&=\sum_{n=-N}^{N}c_n\exp(in\omega t),\\
    z_2^*(t)&=z_1^*(t+T/2)
    =\sum_{n=-N}^{N}(-1)^n c_n\exp(in\omega t),
    \label{SM:eq_AP_fourier}
\end{align}
where $T=2\pi/\omega$.
The Fourier coefficients $c_n$ and angular frequency $\omega$ were determined separately for each periodic solution.
Because $z_k=x_k-i v_k$ is complex-valued, its Fourier coefficients do not in general satisfy the conjugate-symmetry condition $c_{-n}=\bar c_n$.

The Fourier coefficients and frequency were obtained by minimizing the time-averaged residual
\begin{align}
    F_{\rm cost}^{(3)}
    ={}&\frac{\omega}{2\pi}
    \int_0^{2\pi/\omega}
    \left|
    \sum_{n=-N}^{N}in\omega c_n\exp(in\omega t)
    -G_1'\!\left(z_1^*(t),z_2^*(t)\right)
    \right|^2\dd t.
    \label{SM:eq_periodic_cost}
\end{align}
The arbitrary time origin was fixed by setting $\Im c_1=0$.
Fourier modes with $N=20$ were retained.
The integral in Eq.~\eqref{SM:eq_periodic_cost} was evaluated by the trapezoidal rule using 2000 equal subdivisions of one period.

The Fourier coefficients and frequency were estimated using SciPy's \texttt{differential\_evolution} routine with
\begin{align}
    \texttt{tol}=10^{-10},\qquad
    \texttt{atol}=10^{-16},\qquad
    \texttt{maxiter}=5000,\qquad
    \texttt{popsize}=2,\qquad
    \texttt{seed}=0.
\end{align}
The remaining options were set to the SciPy defaults.
We searched
$\Re c_n,\Im c_n\in[-0.5,0.5]$ for odd $n\neq1$,
$\Re c_n,\Im c_n\in[-10^{-6},10^{-6}]$ for even $n$,
$\Re c_1\in[0.1,2]$, and
$\omega\in[0.9,1.3]$.
The actual differential-evolution population contains \texttt{popsize} times the number of free optimization variables.

\subsection{Numerical evaluation of the Floquet exponents \label{SM:numerical_evalualtion_Floquet}}

We computed the Floquet multipliers by a finite-difference approximation of the monodromy matrix.
Let
\begin{align}
    \bm X^*(t)
    =\left(
    \Re z_1^*(t),\Im z_1^*(t),
    \Re z_2^*(t),\Im z_2^*(t)
    \right)^{\mathsf T}
\end{align}
denote either the IP or AP periodic solution.
The unperturbed solution and four solutions initialized with perturbations along the coordinate directions were integrated over one period:
\begin{align}
    &(z_1^*+\varepsilon,z_2^*),\qquad
    (z_1^*+i\varepsilon,z_2^*),\\
    &(z_1^*,z_2^*+\varepsilon),\qquad
    (z_1^*,z_2^*+i\varepsilon).
\end{align}
The perturbation amplitude was set to
\begin{align}
    \varepsilon=5.6\times10^{-3},
\end{align}
which was chosen to balance finite-perturbation and numerical-roundoff errors.
The time evolution was calculated using the Euler method with
\begin{align}
    \Delta t=10^{-4}T,
\end{align}
corresponding to $10^4$ steps per period.

Let $z_k^{(j)}(T)$ denote the endpoint of the $j$th perturbed trajectory.
We define
\begin{align}
    \delta z_k^{(j)}(T)=z_k^{(j)}(T)-z_k^*(0).
\end{align}
The finite-difference monodromy matrix is then
\begin{align}
    M
    =\frac{1}{\varepsilon}
    \begin{pmatrix}
    \Re\delta z_1^{(1)} & \Re\delta z_1^{(2)} & \Re\delta z_1^{(3)} & \Re\delta z_1^{(4)}\\
    \Im\delta z_1^{(1)} & \Im\delta z_1^{(2)} & \Im\delta z_1^{(3)} & \Im\delta z_1^{(4)}\\
    \Re\delta z_2^{(1)} & \Re\delta z_2^{(2)} & \Re\delta z_2^{(3)} & \Re\delta z_2^{(4)}\\
    \Im\delta z_2^{(1)} & \Im\delta z_2^{(2)} & \Im\delta z_2^{(3)} & \Im\delta z_2^{(4)}
    \end{pmatrix}_{t=T}.
    \label{SM:eq_monodromy_coupled}
\end{align}
The Floquet multipliers $\rho_j$ are the eigenvalues of $M$, and the Floquet exponents are evaluated as
\begin{align}
    \Lambda_j=\frac{1}{T}\log\rho_j.
    \label{SM:eq_floquet_exponents}
\end{align}
The imaginary parts are evaluated using the principal value of $\arg\rho_j$.
One Floquet exponent is zero up to numerical accuracy because of time-translation invariance.
The remaining nonzero exponents are ordered by decreasing real part, consistently with the notation used in the main text and figure captions.

For an isolated oscillator, the same procedure was applied to the two-dimensional state
\begin{align}
    \bm X^*(t)=\left(\Re z^*(t),\Im z^*(t)\right)^{\mathsf T}.
\end{align}
Using perturbations $z^*(0)+\varepsilon$ and $z^*(0)+i\varepsilon$, the corresponding monodromy matrix is
\begin{align}
    M_{\rm iso}
    =\frac{1}{\varepsilon}
    \begin{pmatrix}
    \Re\delta z^{(1)}(T) & \Re\delta z^{(2)}(T)\\
    \Im\delta z^{(1)}(T) & \Im\delta z^{(2)}(T)
    \end{pmatrix},
    \label{SM:eq_monodromy_isolated}
\end{align}
where $\delta z^{(j)}(T)=z^{(j)}(T)-z^*(0)$.
The isolated-oscillator Floquet exponents were then calculated from Eq.~\eqref{SM:eq_floquet_exponents}.

\subsection{Comparison with the inferred reduced model}

We numerically analyzed the stability of the in-phase (IP) and anti-phase (AP) synchronized periodic solution of the inferred coupled Stuart-Landau equations, Eq.~(\ref{coupledSL}) of the main text, including all
inferred terms.
The IP periodic solution was represented by the truncated Fourier series
\begin{align}
    z_1^*(t) = z_2^*(t) = \sum_{n=-N}^{N} c_n\exp(in\omega t),
\end{align}
and the AP periodic solution was written as
\begin{align}
    z_1^*(t)
    &=
    \sum_{n=-N}^{N}c_n\exp(in\omega t),
    \label{SM:eq_fluid_AP_z1}\\
    z_2^*(t)
    &=
    z_1^*(t+T/2)
    =
    \sum_{n=-N}^{N}(-1)^n c_n\exp(in\omega t),
    \label{SM:eq_fluid_AP_z2}
\end{align}
where $T=2\pi/\omega$.

The Fourier coefficients $c_n$ and angular frequency $\omega$ were determined by minimizing the time-averaged residual
\begin{align}
    F_{\rm cost}^{(4)}
    ={}&
    \frac{\omega}{2\pi}
    \int_0^{2\pi/\omega}
    \left|
    \sum_{n=-N}^{N}
    in\omega c_n\exp(in\omega t)
    -
    G_1\left(z_1^*(t),z_2^*(t)\right)
    \right|^2
    \dd t.
    \label{SM:eq_fluid_periodic_cost}
\end{align}
The arbitrary time origin was fixed by setting $\Im c_1=0$.
Fourier modes up to $N=6$ were retained, and the integral in Eq.~\eqref{SM:eq_fluid_periodic_cost} was evaluated using the trapezoidal rule with 600 equal subdivisions of one period.

The minimization was performed using SciPy's \texttt{differential\_evolution} routine~\cite{scipy2019} with
\begin{align}
    \texttt{tol}&=10^{-10},&
    \texttt{atol}&=10^{-16},&
    \texttt{maxiter}&=1000,&
    \texttt{seed}&=0.
\end{align}
The remaining options were set to the SciPy defaults.
The parameter bounds were
\begin{align}
    \Re c_n,\Im c_n&\in[-0.5,0.5]
    \qquad (n\neq1),\\
    \Re c_1&\in[0.5,1.5],\\
    \omega&\in[0.95,1.6].
\end{align}

The Floquet exponents of the resulting periodic solution were computed using the finite-difference monodromy-matrix procedure described in Sec.~\ref{SM:numerical_evalualtion_Floquet}.
For the present calculation, the perturbation amplitude was set to
\begin{align}
    \varepsilon=10^{-2},
\end{align}
and the inferred equations were integrated using the Euler method with a time step
\begin{align}
    \Delta t=10^{-4}T,
\end{align}
corresponding to $10^4$ time steps per period.
The zero exponent associated with time-translation invariance was excluded, and the remaining Floquet exponents were ordered by decreasing real part and denoted by $\Lambda_{\mathrm{AP},1}$, $\Lambda_{\mathrm{AP},2}$.

Figures~\ref{fig:E} and \ref{fig:F} compare the inferred coefficients, oscillation frequencies, and Floquet exponents with their analytical or numerical reference values.
Figure~\ref{fig:E} shows their dependence on the coupling strengths, whereas Fig.~\ref{fig:F} shows their dependence on the bifurcation parameter $\mu$.
As the coupling strength or $\mu$ increases, some inferred coefficients deviate from their leading-order analytical values.
Nevertheless, the inferred model continues to reproduce the oscillation frequencies and dominant Floquet exponents over the investigated ranges.
This distinction indicates that agreement with the lowest-order asymptotic coefficients and dynamical accuracy of the inferred reduced model should be assessed separately.

\section{Inference of the coupled van der Pol oscillators from a limited number of trajectories}
\label{SM:limit_number_trajectory}

We examine how the number and dynamical coverage of the trajectories used for fitting affect the inference of the coupling parameters.
In the baseline ensemble inference described in the main text, the initial conditions were prepared near the isolated limit cycle using 21 different initial phase differences $\tau^{(0)}/T$. 
The coupling coefficients were inferred simultaneously from the resulting ensemble of transient trajectories. 
Here, we compare this baseline result with inference using only one or two trajectories.

\begin{figure}[tb]
  \includegraphics[width=14cm]{"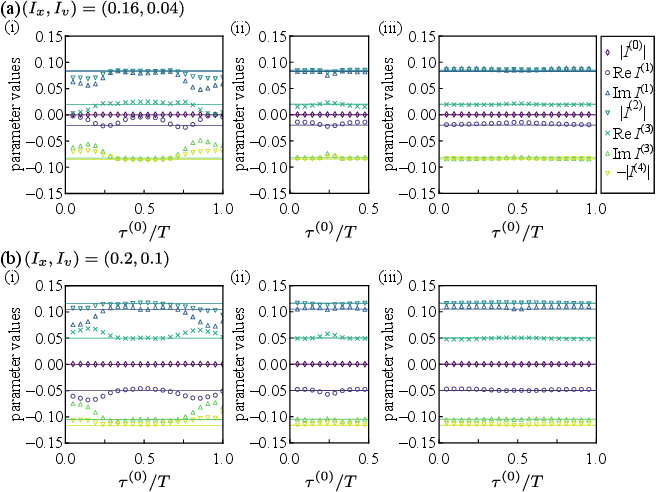"}
  \caption{
  Inference from one or two transient trajectories of the coupled VDP oscillators and validation of the model inferred from a single trajectory.
  (a)--(b) Coupling parameters inferred under different trajectory conditions.
  Results are shown for the bistable regime with stable IP and AP synchronization ($I_x=0.16$ and $I_v=0.04$) and the monostable regime with only stable IP synchronization ($I_x=0.2$ and $I_v=0.1$).
  Panels (i)--(iii) correspond to different trajectory datasets:
  (i) a single trajectory initialized on the isolated limit cycle with initial phase difference $\tau^{(0)}/T$;
  (ii) two trajectories initialized on the isolated limit cycle with phase differences $\tau^{(0)}/T$ and $\tau^{(0)}/T+10/21$, respectively; and
  (iii) a single trajectory initialized with amplitudes substantially smaller than the isolated limit-cycle amplitude, $(x_1,v_1)=(0.1,0)$ and $(x_2,v_2)=(0.1\cos[2\pi\tau^{(0)}/T],-0.1\sin[2\pi\tau^{(0)}/T])$.
  Symbols represent the inferred values, and horizontal lines indicate the reference values.
  }
  \label{fig:D}
\end{figure}

Figure~\ref{fig:D}(a) shows the results in the bistable regime, $I_x=0.16$ and $I_v=0.04$, where both the in-phase (IP) and anti-phase (AP) synchronized states are stable.
Figure~\ref{fig:D}(b) shows the results in the monostable regime, $I_x=0.2$ and $I_v=0.1$, where only the IP state is stable.
In all calculations, we use the same fitting time window and sampling interval 
as the baseline full-ensemble inference specified in Sec.~\ref{SM:sec_vdp_settings}.
Thus, the results below isolate the effect of the number and type of trajectories without additional tuning of the data window.

In panels (i), the inference is performed using a single trajectory whose initial conditions for both oscillators lie on the isolated limit cycle, with an initial phase difference $\tau^{(0)}/T$.
The inferred parameters depend substantially on the selected initial phase difference in both the bistable and monostable regimes.
A single trajectory initialized near the limit cycle typically samples only a restricted region of the coupled state space before approaching a synchronized solution.
Consequently, such a trajectory does not always contain sufficient information to identify all coupling coefficients reliably.

In panels (ii), two trajectories are used.
Their initial phase differences are $\tau^{(0)}/T$ and $\tau^{(0)}/T+10/21$, respectively, so that the two initial conditions are separated by approximately half a period.
Combining these two trajectories substantially reduces the dependence of the inferred parameters on $\tau^{(0)}$ and improves their agreement with the reference values.
In the bistable regime, the two trajectories sample dynamically distinct regions associated with the competing IP and AP synchronized states.
More generally, the improvement indicates that combining trajectories with sufficiently different initial phase differences increases the dynamical coverage of the fitting data and improves parameter identifiability.

In panels (iii), the inference is again performed using only one trajectory, but the initial amplitudes are chosen to be substantially smaller than the amplitude of the isolated limit cycle.
Specifically, the initial conditions are
\begin{align}
 (x_1,v_1)&=(0.1,0),\\
 (x_2,v_2)&=
 \left(
 0.1\cos\frac{2\pi\tau^{(0)}}{T},
 -0.1\sin\frac{2\pi\tau^{(0)}}{T}
 \right).
\end{align}
The resulting trajectory contains a pronounced amplitude-relaxation process before approaching a synchronized state and therefore samples a broader region of state space than a trajectory initialized near the limit cycle.
For both the bistable and monostable regimes, the inferred parameters show close agreement with the reference values over all tested initial phase differences.
Thus, a single trajectory can provide sufficient information for accurate inference when it contains substantial transient amplitude dynamics.

We further examined whether a model inferred from a single trajectory can reproduce the dynamics for initial conditions that were not used in the fitting procedure.
For this test, we used the parameters inferred by the procedure in panel (iii) from the single trajectory with $\tau^{(0)}/T=0.25$.
Without any further parameter fitting, we numerically integrated the resulting coupled SL equations over the same set of initial conditions used to characterize the original coupled VDP system.

The results are shown in Figs.~\ref{fig:2}(b)--(d), which correspond to Figs.~\ref{fig:4}(a)--(c), respectively.
The dynamical structures generated by the inferred SL model closely agree with those of the original VDP system.
In particular, the inferred model reproduces the synchronized states and basin structure over a broad set of initial conditions.
Importantly, only the trajectory with $\tau^{(0)}/T=0.25$ was used for parameter inference; the remaining initial conditions represented in Figs.~\ref{fig:2}(c) and (d) were not included in the fitting data.
This agreement shows that the inferred model reproduces the dynamics beyond the single trajectory used for parameter inference.

These results demonstrate that successful inference does not depend only on the number of trajectories.
The dynamical coverage of the fitting data is also essential.
A single trajectory initialized near the isolated limit cycle may provide insufficient information, whereas two trajectories with substantially different initial phase differences, or one trajectory containing a large amplitude transient, can yield stable parameter estimates.
Moreover, the SL model inferred from a single informative trajectory reproduces the dynamical structures of the original VDP system for initial conditions that were not used for fitting.
This result indicates that the method reconstructs an effective reduced vector field rather than merely reproducing the training trajectory.

Note that all parameter-inference results in Fig.~\ref{fig:D} were obtained using the same time window as the full-ensemble inference in Fig.~\ref{fig:S1} of the main text.
The window was therefore not optimized separately for the one- and two-trajectory datasets.
Because the appropriate fitting window depends on the relaxation dynamics sampled by the selected trajectory, further improvement may be possible by applying the data-window stability analysis described in Sec.~\ref{SM:sec1_window} separately to each limited dataset.

\section{Inference of the three coupled van der Pol oscillators \label{SM:three_coupled}}

\begin{figure}[tb]
  \includegraphics{"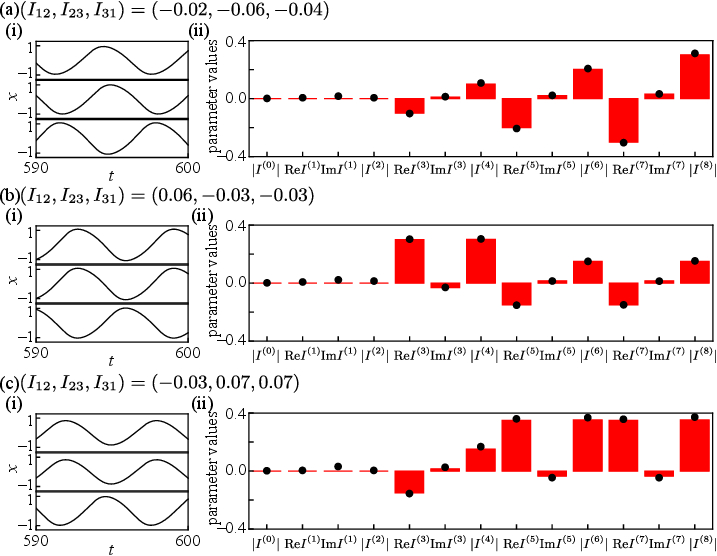"}
  \caption{
  Discrimination by interaction inference for three coupled oscillators under interactions that stabilize partial in‑phase synchronization. 
  The coupling strengths \((I_{12},I_{23},I_{31})\) were set to (a) \((-0.02,-0.06,-0.04)\), (b) \((0.06,-0.03,-0.03)\), and (c) \((-0.03,0.07,0.07)\). 
  Panels: 
  (i) representative waveforms of the partially in‑phase synchronized state; 
  (ii) inferred coupling parameters (markers) and theoretical values (bar graphs).
  \label{fig:G}}
\end{figure}

We performed the numerical calculation for 
\begin{gather}
    \ddot x_1=(\mu-x_1^2)\dot x_1-x_1+I_{12}\dot x_2+I_{31}\dot x_3\\
    \ddot x_2=(\mu-x_2^2)\dot x_2-x_2+I_{12}\dot x_1+I_{23}\dot x_3\\
    \ddot x_3=(\mu-x_3^2)\dot x_3-x_3+I_{31}\dot x_1+I_{23}\dot x_2
\end{gather}
until $t=0$ to ensure the convergence to each limit cycle, where we set the initial conditions $(x_1,\dot x_1)=(0.1,0)$ at $t=-300$, $(x_2,\dot{x}_2)=(0.1,0)$ at $t=-300+\tau^{(0)}$, and $(x_3,\dot{x}_3)=(0.1,0)$ at $t=-300+\tilde\tau^{(0)}$. $\tau^{(0)}$ and $\tilde\tau^{(0)}$ are the initial time delays among the three oscillators.
We changed the coupling strength from $0$ to the appropriate values at $t=0$ and continued the numerical calculations until $t=300$.
We set $\mu=0.2$ and varied $I_x$ and $I_v$ in the ranges $0\leq I_v\leq 0.1$ and $0\leq I_x\leq 0.2$. $\tau^{(0)}$ was varied from $T/7$ to $6T/7$ in increments of $5T/21$ and $\tilde\tau^{(0)}$ was varied from $4T/21$ to $19T/21$ in increments of $5T/21$, where $T$ is the intrinsic period of the single oscillator. 

We assumed the form of the three-coupled Stuart-Landau equations as
\begin{align}
    \dot z_1=G_1(z_1,z_2,z_3)=(a+ib)z_1-(a+id)|z_1|^2z_1+I^{(0)}+I^{(1)}z_1+I^{(2)}\bar z_1+I^{(3)}z_2+I^{(4)}\bar z_2+I^{(5)}z_3+I^{(6)}\bar z_3\\
    \dot z_2=G_2(z_2,z_1,z_3)=(a+ib)z_2-(a+id)|z_2|^2z_2+I^{(0)}+I^{(1)}z_2+I^{(2)}\bar z_2+I^{(3)}z_1+I^{(4)}\bar z_1+I^{(7)}z_3+I^{(8)}\bar z_3\\
    \dot z_3=G_3(z_3,z_1,z_2)=(a+ib)z_3-(a+id)|z_3|^2z_3+I^{(0)}+I^{(1)}z_3+I^{(2)}\bar z_3+I^{(5)}z_1+I^{(6)}\bar z_1+I^{(7)}z_2+I^{(8)}\bar z_2
\end{align}
and determined the coefficients by minimizing
\begin{align}
    F_{\rm cost}^\qty(5)=\sum_{\rm i.c.}\frac{1}{t_{\rm w}}\int_{t_{\rm s}}^{t_{\rm s}+t_{\rm w}}\qty(\qty|\dot{\hat z}_1-G_1(\hat z_1,\hat z_2,\hat z_3)|^2+\qty|\dot{\hat z}_2-G_2(\hat z_2,\hat z_1,\hat z_3)|^2+\qty|\dot{\hat z}_3-G_3(\hat z_3,\hat z_1,\hat z_2)|^2)\dd t.
\end{align}
The conditions for the time series used in the inference were the same as in the two-oscillator case.

As shown in Fig.~\ref{fig:G}, the inferred parameters are in excellent agreement with the theoretical values.

\section{Inference of the coupled van der Pol oscillators under noise \label{SM:noisy_system}}

\begin{figure}[tb]
  \includegraphics{"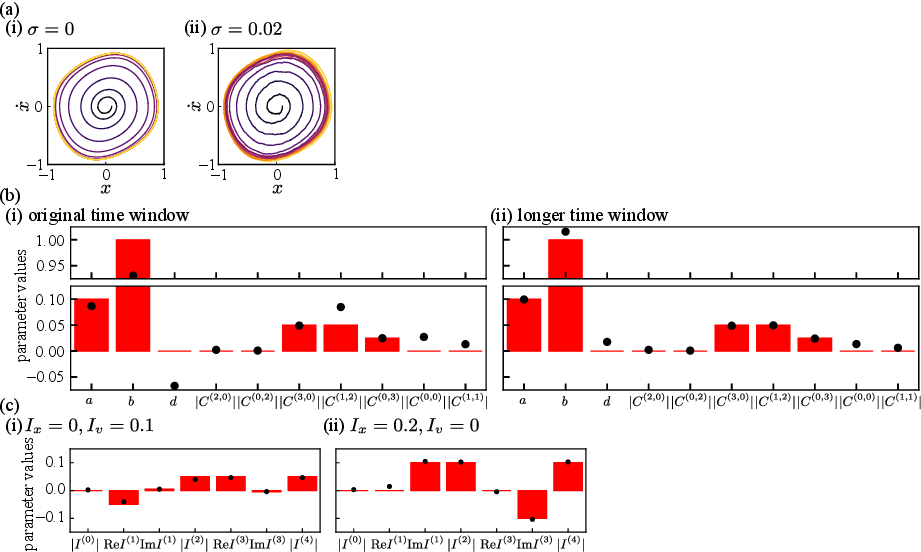"}
  \caption{
    Robustness of the inference under noise. (a) Trajectories in the \(x\)-\(\dot x\) phase space for (i) noise intensity $\sigma=0$ and (ii) $\sigma=0.02$.
  (b) Single-oscillator inference: (i) results obtained with the same time window as in the noiseless case, and (ii) results obtained with a longer time window.
  (c) Inference results for the coupled system.
  In panels (b) and (c), inferred parameter values are shown by markers and theoretical values by bar graphs.
  \label{fig:H}}
\end{figure}

The single oscillator with additive white noise was modeled as
\begin{align}
    \ddot x=(\mu-x^2)\dot x-x+\xi,
\end{align}
where $\xi$ denotes zero-mean white noise with $\langle\xi(t)\xi(t')\rangle = \sigma^2 \delta(t-t')$.
For the coupled system we considered
\begin{align}
    \ddot x_k=(\mu-x_k^2)\dot x_k-x_k-I_x(x_k-x_l)-I_v(\dot x_k-\dot x_l)+\xi_k,
\end{align}
where $\langle\xi_k(t)\xi_\ell(t')\rangle = \sigma^2\delta_{k\ell}\delta(t-t')$.
Numerical integration was performed using the Euler-Maruyama method.

We integrated the system with $I_x=I_v=0$ up to $t=0$ to ensure convergence to the limit cycles, setting initial conditions $(x_1,\dot x_1)=(0.1,0)$ at $t=-300$ and $(x_2,\dot{x}_2)=(0.1,0)$ at $t=-300+\tau^{(0)}$, where $\tau^{(0)}$ is the initial delay between the two oscillators.
At $t=0$ the coupling strengths were switched to the target values and integration continued until $t=300$.
We fixed $\mu=0.2$, varied $I_v\in[0,0.1]$ and $I_x\in[0,0.2]$, and sampled $\tau^{(0)}$ from $0$ to $T$ in steps of $T/21$, where $T$ is the intrinsic period of the single oscillator.

Using the time-series conditions employed for the noiseless case, we first inferred the single-oscillator parameters.
Figure~\ref{fig:H}(b)(i) shows the result obtained with the original (narrow) time-window settings; a noticeable discrepancy is observed. 
To reduce this discrepancy, we enlarged the time window: Fourier coefficients were computed from $t\in[-16T,0]$ with $\Delta t=0.01$, and the single Stuart-Landau parameters were inferred from data on $t\in[-300,-275]$ with $\Delta t=0.1$.
The resulting fit, shown in Fig.~\ref{fig:H}(b)(ii), exhibits a reduced discrepancy.

For the coupled system, inference using the same (noiseless-case) time-series conditions produced the results shown in Fig.~\ref{fig:H}(c); no significant issues were observed.

\section{Numerical Floquet analysis of the inferred fluid-system model}
\label{SM:Fluid_system_Floquet}

We numerically analyzed the stability of the anti-phase (AP) synchronized periodic solution of the inferred coupled Stuart-Landau equations, Eq.~(\ref{coupledSL}) of the main text, including all inferred terms.
The AP periodic solution was represented by the truncated Fourier series
\begin{align}
    z_1^*(t)
    &=
    \sum_{n=-N}^{N}c_n\exp(in\omega t),
    \label{SM:eq_fluid_AP_z1a}\\
    z_2^*(t)
    &=
    z_1^*(t+T/2)
    =
    \sum_{n=-N}^{N}(-1)^n c_n\exp(in\omega t),
    \label{SM:eq_fluid_AP_z2a}
\end{align}
where $T=2\pi/\omega$.

The Fourier coefficients $c_n$ and angular frequency $\omega$ were determined by minimizing the time-averaged residual
\begin{align}
    F_{\rm cost}^{(6)}
    ={}&
    \frac{\omega}{2\pi}
    \int_0^{2\pi/\omega}
    \left|
    \sum_{n=-N}^{N}
    in\omega c_n\exp(in\omega t)
    -
    G_1\left(z_1^*(t),z_2^*(t)\right)
    \right|^2
    \dd t.
    \label{SM:eq_fluid_periodic_cost2}
\end{align}
The arbitrary time origin was fixed by setting $\Im c_1=0$.
Fourier modes up to $N=6$ were retained, and the integral in
Eq.~\eqref{SM:eq_fluid_periodic_cost2} was evaluated using the trapezoidal rule with 600 equal subdivisions of one period.

The minimization was performed using SciPy's \texttt{differential\_evolution} routine~\cite{scipy2019} with
\begin{align}
    \texttt{tol}&=10^{-10},&
    \texttt{atol}&=10^{-16},&
    \texttt{maxiter}&=2000,&
    \texttt{seed}&=0.
\end{align}
The remaining options were set to the SciPy defaults.
The parameter bounds were
\begin{align}
    \Re c_n,\Im c_n&\in[-0.5,0.5]
    \qquad (n\neq1),\\
    \Re c_1&\in[0.1,1.1],\\
    \omega&\in[2,3].
\end{align}

The Floquet exponents of the resulting periodic solution were computed using the finite-difference monodromy-matrix procedure described in Sec.~\ref{SM:numerical_evalualtion_Floquet}.
For the present calculation, the perturbation amplitude was set to
\begin{align}
    \varepsilon=10^{-2},
\end{align}
and the inferred equations were integrated using the Euler method with a time step
\begin{align}
    \Delta t=10^{-4}T,
\end{align}
corresponding to $10^4$ time steps per period.
The zero exponent associated with time-translation invariance was excluded, and the remaining Floquet exponents were ordered by decreasing real part and denoted by
$\Lambda_{\mathrm{AP},1}$ and $\Lambda_{\mathrm{AP},2}$.

The resulting Floquet exponents are shown in Fig.~\ref{fig:3} of the main text.
The real parts of the complex-conjugate pair $\Lambda_{\mathrm{AP},1}$ and $\Lambda_{\mathrm{AP},2}$ cross zero near $\delta\simeq \delta_{\rm c}$, while their imaginary parts remain nonzero.
This identifies the loss of AP stability as a Neimark-Sacker
bifurcation.
The predicted bifurcation point closely agrees with the loss of AP synchrony observed in the original hydrodynamic simulations and with our previous report~\cite{Araya2024}.

The above analysis characterizes the local stability of the AP periodic solution.
In the following section, we examine whether the inferred model can also reproduce the global basin structure of the competing IP and AP synchronized states when only two transient fluid-system trajectories are used for parameter inference.

\section{Reconstruction of the basin structure from two fluid-system trajectories}
\label{SM:sec_fluid_basin_limited}

\begin{figure}[tb]
  \includegraphics[width=14cm]{"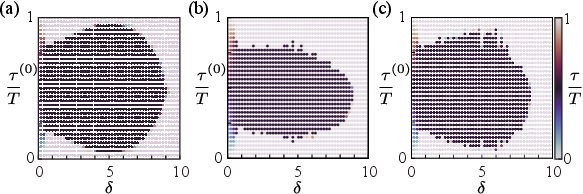"}
  \caption{
  Basin structure of the coupled oscillatory flows reconstructed from two transient trajectories.
  The color indicates the final phase difference $\tau/T$ after transients in the plane spanned by the initial phase difference $\tau^{(0)}/T$ and the coupling parameter $\delta$.
  (a) Reference result obtained from the original hydrodynamic simulations.
  (b)--(c) Results obtained by numerically integrating the inferred coupled Stuart-Landau equations.
  For each coupling condition, the inference used only two transient trajectories selected so that, in the bistable regime, one trajectory converged to the in-phase synchronized state and the other converged to the anti-phase synchronized state.
  The fitting windows started at (b) $t_{\rm s}=80$ and (c) $t_{\rm s}=0$.
  Initial conditions used to evaluate the remaining points in the basin maps were not included in the parameter inference.
  }
  \label{fig:fluid_basin_limited}
\end{figure}

We examined whether the basin structure of the hydrodynamic system can be reconstructed from a limited number of transient trajectories.
In the inference used for the main hydrodynamic results, waveform data were collected from multiple initial phase differences.
Here, we restricted the fitting data to two transient trajectories for each coupling condition.
The two initial conditions were selected so that, in the bistable regime, one trajectory converged to the in-phase (IP) synchronized state and the other converged to the anti-phase (AP) synchronized state.
Thus, the two trajectories sampled dynamically distinct regions associated with the competing synchronized states.

For each inferred coupled Stuart-Landau (SL) model, we numerically integrated the inferred equations from a grid of initial phase differences $\tau^{(0)}/T$ and coupling strengths $\delta$.
After discarding the transient, we calculated the final phase difference $\tau/T$ and compared the resulting basin map with that obtained directly from the hydrodynamic simulations.
The initial conditions used to construct the basin maps were not included in the parameter inference, except for the two trajectories described above.

Figure~\ref{fig:fluid_basin_limited}(a) shows the reference basin structure obtained from the original hydrodynamic simulations.
Figures~\ref{fig:fluid_basin_limited}(b) and \ref{fig:fluid_basin_limited}(c) show the corresponding results obtained from the inferred coupled SL equations using fitting windows starting at $t_{\rm s}=80$ and $t_{\rm s}=0$, respectively.
In both cases, the inferred model captures the coexistence of the IP and AP synchronized states and the main dependence of their basins on the initial phase difference and coupling strength.
The agreement is not exact, especially near the basin boundary and near the coupling range where the AP basin disappears.
Nevertheless, the main basin structure is recovered from only two transient trajectories.

The result obtained with $t_{\rm s}=0$ more closely reproduces the overall shape of the basin structure than the result obtained with $t_{\rm s}=80$.
The earlier fitting window contains a larger portion of the relaxation dynamics and therefore samples a broader region of the phase-amplitude state space.
By contrast, the later window contains trajectories closer to the final synchronized states and provides less information about the transient organization of the two basins.
This comparison supports the conclusion that the dynamical coverage of the fitting data, rather than the number of trajectories alone, is important for reconstructing multistable dynamics.

The two trajectories and the two fitting windows used here were selected as a focused limited-data test and were not optimized separately to maximize agreement of the basin maps.
The remaining discrepancies may therefore reflect the limited state-space coverage of the selected trajectories and the use of a common fitting-window protocol.
Further improvement may be possible by selecting trajectories with richer amplitude-relaxation dynamics and by applying the data-window stability analysis in Sec.~\ref{SM:sec1_window} specifically to the limited fluid-system dataset.

\end{document}